\documentclass[reprint,aps,showpacs,nofootinbib]{revtex4-1}

\usepackage{ulem} 
\usepackage{color}

\definecolor{green}{rgb}{0,.5,0}

\definecolor{red}{rgb}{1,0,0}

\usepackage{graphicx}
\usepackage{subfig}
\usepackage{placeins}
\usepackage{epsfig}
\usepackage{dcolumn}
\usepackage{amsmath}
\usepackage{rotating}
\usepackage{latexsym}
\usepackage{multirow}
\usepackage{mathrsfs} 
\usepackage{slashed}
\usepackage{bbold}
\usepackage[format=plain,singlelinecheck=false,justification=RaggedRight]{caption}

\newcommand{\RNum}[1]{\uppercase\expandafter{\romannumeral #1\relax}}

\def\be{\begin{equation}}
\def\ee{\end{equation}}
\def\bea{\begin{eqnarray}}
\def\eea{\end{eqnarray}}

\begin{document}

\title{\boldmath Trace anomaly form factors from lattice QCD}

\author{Bigeng Wang$^{1,2}$, Fangcheng He$^{3}$, Gen Wang$^4$, Terrence Draper$^1$, Jian Liang$^{5,6}$, Keh-Fei Liu$^{1,2}$, Yi-Bo Yang$^{7,8,9,10}$
\vspace*{-0.5cm}
\begin{center}
\large{
\vspace*{0.4cm}
\includegraphics[scale=0.15]{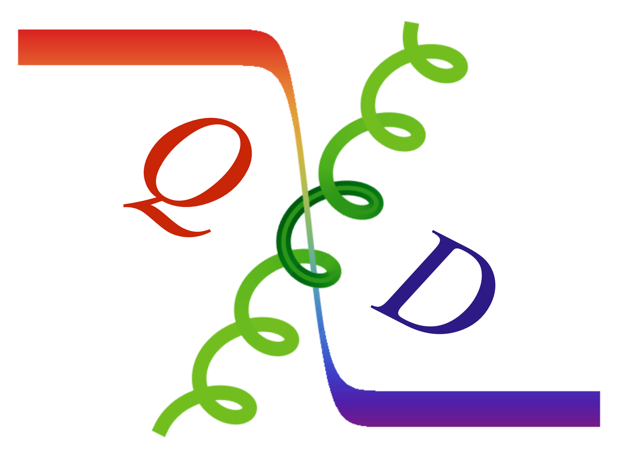}\\
\vspace*{0.4cm}
($\chi$QCD Collaboration)
}
\end{center}
}
\affiliation{
$^{1}$\mbox{Department of Physics and Astronomy, University of Kentucky, Lexington, Kentucky, USA}\\
$^{2}$\mbox{Nuclear Science Division, Lawrence Berkeley National Laboratory, Berkeley, California 94720, USA} \\
$^{3}$\mbox{Department of Physics and Astronomy, Stony Brook University, Stony Brook, New York 11794, USA}\\
$^{4}$\mbox{Aix-Marseille Université, Université de Toulon, CNRS, CPT, Marseille UMR 7332, France}\\
$^{5}$\mbox{Key Laboratory of Atomic and Subatomic Structure and Quantum Control (MOE),}
      \mbox{Guangdong Basic Research Center of Excellence for Structure and Fundamental Interactions of Matter,}
      \mbox{Institute of Quantum Matter, South China Normal University, Guangzhou 510006, China}
$^{6}$\mbox{Guangdong-Hong Kong Joint Laboratory of Quantum Matter,}
       \mbox{Guangdong Provincial Key Laboratory of Nuclear Science,} 
       \mbox{Southern Nuclear Science Computing Center, South China Normal University, Guangzhou 510006, China}\\
$^{7}$\mbox{CAS Key Laboratory of Theoretical Physics, Institute of Theoretical Physics,} \\
      \mbox{Chinese Academy of Sciences, Beijing 100190, China}\\
$^{8}$\mbox{School of Fundamental Physics and Mathematical Sciences,} \\
      \mbox{Hangzhou Institute for Advanced Study, UCAS, Hangzhou 310024, China}\\
$^{9}$\mbox{International Centre for Theoretical Physics Asia-Pacific, Beijing/Hangzhou, China}\\
$^{10}$\mbox{School of Physical Sciences, University of Chinese Academy of Sciences, Beijing 100049, China}\\
}

\begin{abstract}
The hadron mass can be obtained through the calculation of the trace of the energy-momentum tensor in the hadron which includes the trace anomaly and sigma terms. The anomaly due to conformal symmetry breaking is believed to be an important ingredient for hadron mass generation and confinement. In this work, we will present the calculation of the glue part of the trace anomaly form factors of the pion up to $Q^2\sim 4.3~\mathrm{GeV}^2$ and the nucleon up to $Q^2\sim 1~\mathrm{GeV}^2$. The calculations are performed on a domain wall fermion ensemble with overlap valence quarks at seven valence pion masses varying from $\sim 250$ to $\sim 540$ MeV, including the unitary point $\sim 340$ MeV. We calculate the radius of the glue trace anomaly for the pion and the nucleon from the $z$ expansion. By performing a two-dimensional Fourier transform on the glue trace anomaly form factors in the infinite momentum frame with no energy transfer, we also obtain their spatial distributions for several valence quark masses. The results are qualitatively extrapolated to the physical valence pion mass with systematic errors from the unphysical sea quark mass, discretization effects in the renormalization sum rule, and finite-volume effects to be addressed in the future.  We find the pion's form factor changes sign, as does its spatial distribution, for light quark masses. This explains how the trace anomaly contribution to the pion mass approaches zero toward the chiral limit.
\end{abstract}

\maketitle
\section{Introduction}
In classical physics, scale symmetry is broken by the mass term. However, when quantum effects are considered, this scale symmetry can be further broken, leading to the scale anomaly, corresponding to the anomaly which appears in the trace of energy-momentum tensor (EMT). The quantum chromodynamics (QCD) trace anomaly is very important in understanding confinement since it supplies a constant negative pressure which 
cancels the positive pressure from the quarks and glue for the confined hadron in equilibrium~\cite{Liu:2021gco_PhysRevD.104.076010,liu2023hadrons}. Also, the trace anomaly plays an important role in hadron mass generation in a QCD system. The hadron mass can be obtained through the trace term of the EMT ~\cite{Collions_MS_Trace_anomaly_PhysRevD.16.438,TARRACH_Renorm_FF_198245,Shifman:1978zn,Ji_T00_PhysRevLett.74.1071} and the sigma terms,
\begin{equation}
                m_{\mathrm{H}} =   \langle \frac{\beta}{2g}F^2 + \sum_f \gamma_m m_f \overline{\psi}_f \psi_f \rangle_{\mathrm{H}}
                +\sum_f m_f \langle\overline{\psi}_f \psi_f  \rangle_{\mathrm{H}} \label{eqn:hadron_mass_TA}
\end{equation}
where $\langle O\rangle_{\mathrm{H}}\equiv \langle \mathrm{H}| \int \mathrm{d}^3x\,\gamma\,O(x)|\mathrm{H}\rangle/\langle \mathrm{H}|\mathrm{H}\rangle$ is the normalized matrix element of the operator $O$ in the rest frame, where $\mathrm{H}$ denotes the hadron of interest and $m_f$ is the quark mass for the $f$ flavor. The bracketed first term on the right-hand side is the trace anomaly term, and the second is the sigma term. They are separately renormalization group invariant. One can obtain their contributions to the mass of the different hadron states using the above equation~\cite{Yang:2018nqn}. For the nucleon, the sigma term is small (i.e., $\sim$ 8.5\% of the nucleon mass~\cite{Liu:2021gco_PhysRevD.104.076010,liu2023hadrons}), and thus the glue part of the trace anomaly dominates. For the pion, the trace anomaly term contributes about half of the pion mass and, since $\gamma_m \sim 0.2$ is not large, the glue part dominates the trace anomaly term. Since these are nonperturbative quantities, lattice QCD calculation is indispensable for obtaining results with controlled statistical and systematic errors.

The measurement of the trace anomaly form factor and understanding its role in the hadron mass are of great interest and are considered to be one of the major scientific goals of the Electron-Ion Collider \cite{EIC_trace_anomaly}. 

In addition to the nucleon trace anomaly form factor, that of the pion might be more interesting and intriguing. It is 
pointed out that, in view of the fact that the pion sigma term in Eq.~(\ref{eqn:hadron_mass_TA}) gives half of the pion mass from the Gellmann-Oakes-Renner relation and the the Feynman-Hellman theorem, the trace anomaly takes the other half of the pion mass and, therefore, the glue part of the pion trace anomaly will also be proportional to $\sqrt{m}$, just as are the pion mass and its sigma term~\cite{liu2023hadrons}. This poses a puzzle as to why the trace anomaly, which is from the conformal symmetry breaking, should have such a chiral-symmetry-related behavior and what kind of structure change can facilitate this attribute when approaching the chiral limit~\cite{liu2023hadrons}. In light of this puzzle, a lattice calculation has been carried out to examine a spatial distribution in the nucleon, the $\rho$, and the pion~\cite{Fangcheng_PhysRevD.104.074507}, where the spatial coordinate is between the glue part of the trace anomaly operator and the sink position of the interpolation field of the hadron. It was found that the density distributions for the nucleon and the $\rho$ are monotonic as are the electric and axial charge distributions. However, the distribution for the pion is unusual. When the quark mass is small, the distribution changes sign such that the integral of the distribution vanishes at the chiral limit. This is achieved by making the glue trace anomaly more negative than that in the vacuum in the inner core of the pion and more positive than that of the vacuum in the outer shell so that it takes no energy to create a pion with massless quarks.
This finding has motivated the present work to study the glue part of the trace anomaly form factor for the nucleon and particularly the pion. It is predicted that the pion trace anomaly form factor will change sign~\cite{liu2023hadrons} as does the spatial distribution in Ref.~\cite{Fangcheng_PhysRevD.104.074507}.

This paper is organized as follows: In Sec.~\ref{sec:numerical_setup}, we present the numerical details of this calculation and a brief description of grid source propagators and the low-mode substitution method which enables us to obtain a fairly large number of momentum transfer cases. Fits for the form factors, $z$-expansion fits, and the corresponding glue trace anomaly spatial distributions and mass radii for the hadrons are discussed in Sec.~\ref{sec:results}. A summary and outlook are given in Sec.~\ref{sec:conclusion_and_outlook}.

\section{Numerical Setup}
\label{sec:numerical_setup}
We use overlap fermions on one ensemble of (2+1)-flavor domain-wall fermion configurations with Iwasaki gauge action (labeled with letter I) \cite{DWF_24I_PhysRevD.83.074508} as listed in Table \ref{tab:lattice_info}.

The effective quark propagator of the massive overlap fermions is the inverse of the operator $(D_\mathrm{c} + m)$ \cite{TW_Chiu_Prop_Ginsparg-Wilson_PhysRevD.60.034503,Liu:2002qu}, where $D_\mathrm{c}$ is chiral, i.e., $\{D_\mathrm{c} , \gamma_5 \} = 0$ \cite{TW_Chiu_Sol_Ginsparg-Wilson_PhysRevD.59.074501}. It can be expressed in terms of the overlap Dirac operator $D_{\mathrm{ov}}$ as $D_c = \rho D_{\mathrm{ov}} /(1 - D_{\mathrm{ov}} /2)$, with $\rho = -(1/(2\kappa) - 4)$ and $\kappa$ = 0.2. A multimass inverter is used to calculate the propagators with seven valence pion masses varying from $\sim 250$ to $\sim 540$ MeV, including the unitary point $\sim 340$ MeV. On the 24I ensemble, hypercubic smearing is applied to the gauge links in the overlap fermion and Gaussian smearing \cite{Gaussian_Smearing_DEGRAND199184} is applied with root mean square (rms) radius 0.49 fm \cite{Pion_FF_Gen_Wang:2020nbf}, respectively, for both the source and sink. 

\begin{table}[!htbp]
    \begin{tabular}{cccccccc} \hline \hline
    Ensemble & $L^3\times T$  &$a$ (fm)  & $L (\rm fm)$ &  $m_{\pi}$ (MeV)  & $N_{\mathrm{conf}}$ & $N_{\mathrm{src}}$\\
\hline

    24I & $24^3\times 64$ & 0.1105(3) & 2.65  &340  &  788 & $64 \times 2$\\
    \hline \hline
    \end{tabular}
    \caption{Details of the 24I ensemble used in this calculation. We use the grid source with two sets of the $Z_3$ noise~\cite{Dong:1993pk} for 64 smeared grids on each time slice. 
    }
    \label{tab:lattice_info}
\end{table}

Based on the normalization convention shown in the Appendix, we define a dimensionless mass form factor $\mathcal{F}_{\mathrm{m},\mathrm{H}}(Q^2)$, where $Q^2 = -(p'-p)^2$. For a spin-$\frac{1}{2}$ particle like the nucleon, we have
\begin{equation}
    \label{eqn:nucleon_TAFF}
    \langle p',\textbf{s}' | T^{\mu}_{\mu}| p , \textbf{s} \rangle = m_{\mathrm{N}}\mathcal{F}_{\mathrm{m},\mathrm{N}}(Q^2)\bar{u}(p',\textbf{s}') u(p,\textbf{s}),
\end{equation}
where $T^{\mu}_{\mu}$ is the trace operator of the EMT
\begin{align}
    \label{trace}
    \begin{aligned}
        T^{\mu}_{\mu} = 
        \frac{\beta}{2g}F^2 + \sum_f (1+\gamma_m) m_f \overline{\psi}_f \psi_f,
    \end{aligned}
\end{align}
and $\textbf{s}$ ($\textbf{s}'$) is the canonical polarization of the initial (final) nucleon.

For a spin-0 particle like the pion, we have
\begin{equation}
\label{eqn:pion_TAFF}
\langle p' | T^{\mu}_{\mu}| p \rangle = m_{\pi}\,\mathcal{F}_{\mathrm{m},\pi}(Q^2).
\end{equation}

The trace anomaly operator $T^{\mu}_{\mu}$ is composed of two parts
\begin{align}
     T^{\mu}_{\mu} &= (T^{\mu}_{\mu})_{\sigma} + (T^{\mu}_{\mu})_{a},
 \end{align}
where $(T^{\mu}_{\mu})_{\sigma} = \sum_{f} m_f \overline{\psi}_f \psi_f$ and $(T^{\mu}_{\mu})_{a} = \Big [ \sum_{f} m_f \gamma_m(g) \overline{\psi}_f \psi_f + \frac{\beta(g)}{2g}F^2 \Big ]$.
And correspondingly the form factor of the EMT trace of the hadron $\mathrm{H}$, i.e., the mass form factor $\mathcal{F}_{\mathrm{m},\mathrm{H}}(Q^2)$, is made up
of two parts:
\begin{equation} \label{G_m}
\mathcal{F}_{\mathrm{m},\mathrm{H}} (Q^2) = \mathcal{F}_{\rm{ta},\mathrm{H}} (Q^2)+ \mathcal{F}_{\sigma,\mathrm{H}} (Q^2),
\end{equation}
where $\mathcal{F}_{\sigma,\mathrm{H}}$ is the form factor of the sigma term and $\mathcal{F}_{\rm{ta},\mathrm{H}}$ is the form factor of the trace anomaly. In the forward limit where $Q^2=0$, $\mathcal{F}_{\mathrm{m},\mathrm{H}} (Q^2=0) = 1 $. In this work, we calculate the form factor of the glue part of the trace anomaly i.e.,  $G_{\mathrm{H}} \equiv \frac{\beta(g)}{2g}\langle F^2
\rangle_\mathrm{H}/m_{\mathrm{H}}$ for the nucleon and the pion. 

\begin{figure}[htbp]
    
    \includegraphics[width=0.45\textwidth]{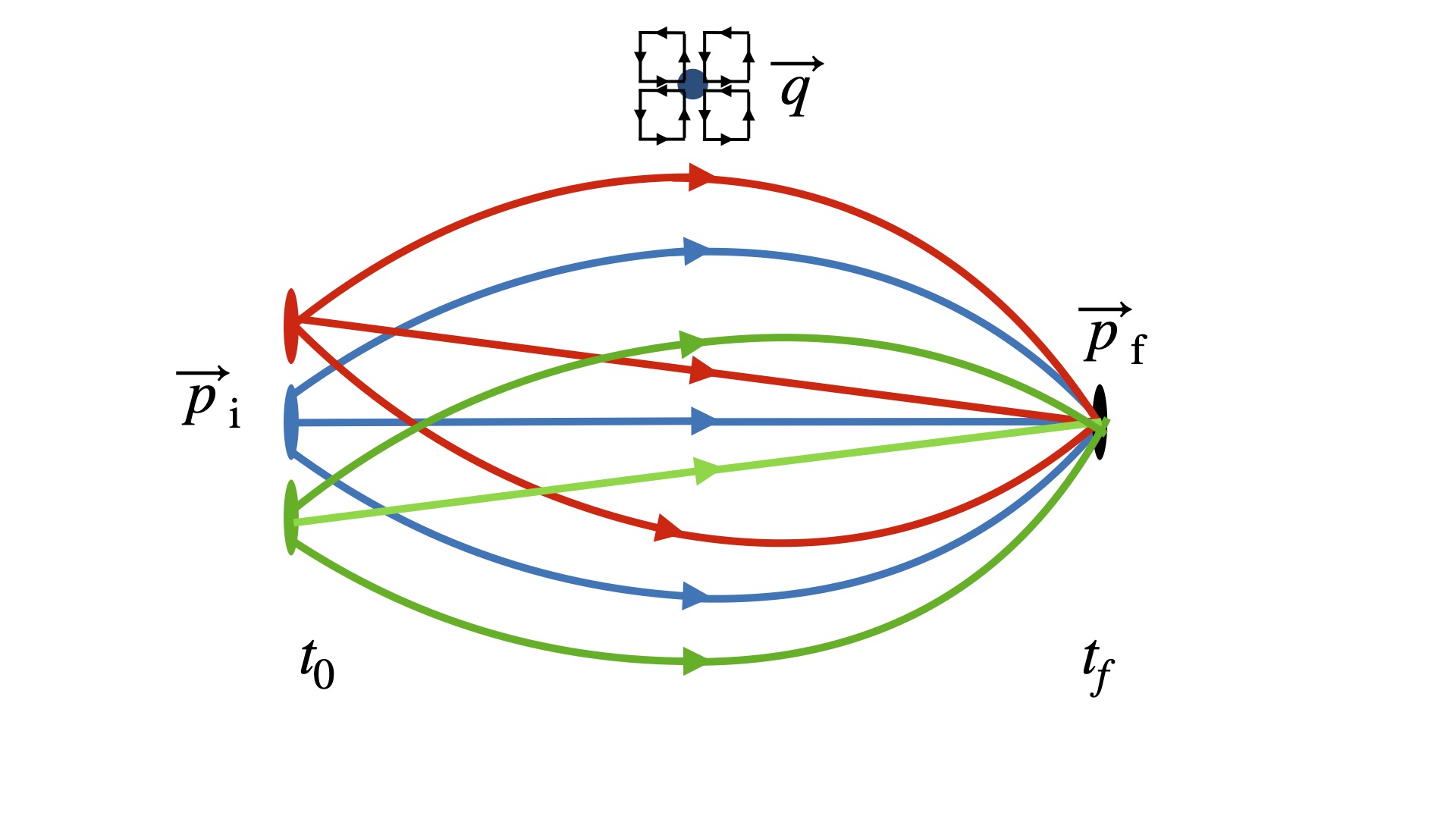}
    \caption{The contraction of three-point function from two-point hadron propagators and glue operators. The different colors represent different contributions from the grid points of the source.}
    \label{fig:grid_source_glue_operator}
\end{figure}

To extract matrix elements and form factors, the three-point function (3pt) $C_3(\vec{q};t_f,\tau;t_0)$ is computed from the contraction of the two-point hadron propagator and the glue operator $F^2$, which is built from the ``cloverleaf" link operators with the hypercubic~\cite{Hasenfratz:2001hp} smeared gauge links \cite{Gen_Proton_Mom_PhysRevD.106.014512}.  Using the nucleon as an example, the contraction is shown in Fig. \ref{fig:grid_source_glue_operator}.

We use the low-mode substitution method \cite{xQCD:2010pnl, XQCD:2013odc, Yang:2015zja, Liang:2016fgy, Liang:2018pis, Pion_FF_Gen_Wang:2020nbf,Gen_Proton_Mom_PhysRevD.106.014512} with grid-source propagators with $Z_3$ noises \cite{Dong:1993pk}. The sink momentum $\vec{p}_f$ is applied after the contraction, and the source momentum $\vec{p}_i$ is applied by multiplying phase factors to the grid points together with the low modes in the low-mode substitution and corrected with a point source propagator for the high modes~\cite{Gen_Proton_Mom_PhysRevD.106.014512}. Compared to the traditional method where the momentum projection for the source is implemented before the inversion for propagators, in this construction, the momentum projection for the source and sink can both be implemented at the correlation function level and can reduce the inversion cost by a factor proportional to the number of source momenta. Therefore we can enhance our data with various momentum transfers in the above-mentioned scenarios, without a significant increase in the computational cost. 

While we would like to get results with the magnitude of momentum transfer $Q^2 = -(p_f - p_i)^2$ as large as possible, larger $\vec{p}_{\mathrm{i}}$ and $\vec{p}_{\mathrm{f}}$ usually lead to worse signals and larger finite lattice spacing errors. Therefore following the choices in  Ref.~\cite{Gen_Proton_Mom_PhysRevD.106.014512}, we choose three momentum transfer scenarios where for a given $Q^2$ we can use as small  $|\vec{p}_{\mathrm{i}}|$ and $|\vec{p}_{\mathrm{f}}|$ as possible. The three momentum transfer scenarios are as follows: (1) The source-at-rest case: $|\vec{p}_{\mathrm{i}}| = 0$ with $\vec{q} = \vec{p}_{\mathrm{f}}$. This will cover the smallest $Q^2$ values where a sign change of the form factor may be found.
(2) The back-to-back case: $\vec{p}_{\mathrm{f}} = - \vec{p}_{\mathrm{i}}$ with $\vec{q} = 2 \vec{p}_{\mathrm{f}} $, which will cover largest $Q^2$ values and show the asymptotic behavior of the form factors.
(3) The near-back-to-back case: for a given $\vec{q}$, $\vec{p}_{\mathrm{f}}$ and $-\vec{p}_{\mathrm{i}}$ are close to $\vec{q}/2$,  which serves as a supplement to the data obtained in the above-mentioned two cases.

As shown in Fig.~\ref{fig:grid_source_glue_operator}, the diagram for the glue trace anomaly form factors is disconnected and we also need to subtract the vacuum expectation values from the 3pt. The subtracted 3pt with various momentum transfers can be written as
\begin{align}
    \begin{aligned}
        \langle C_3(\vec{q};t_f,\tau;t_0) \rangle =& \langle C_{2}(\vec{p}_i, \vec{p}_f,t) \times O(\vec{q},\tau)\rangle \\
        &- \langle C_{2}(\vec{p}_i, \vec{p}_f,t)\rangle \times \langle O(\vec{q},\tau)\rangle \\
        =&\langle C_{2,\mathrm{sub}}(\vec{p}_i, \vec{p}_f,t) \times O_{\mathrm{sub}}(\vec{q},\tau) \rangle,
    \end{aligned}
\end{align}
where we have the hadron propagator with vacuum expectation value subtracted $C_{2,\mathrm{sub}}(\vec{p}_i, \vec{p}_f,t)=   C_2(\vec{p}_i, \vec{p}_f,t)  - \langle C_2(\vec{p}_i, \vec{p}_f,t) \rangle$,
and the glue operator $O=F^2$ with momentum $\vec{q} = \vec{p}_f - \vec{p}_i$ is defined as $O(\vec{q},\tau) = \sum _{\vec{z}} e^{+i \vec{q} \cdot \vec{z} }  O(\vec{z},\tau)$, and we use the glue operator with vacuum expectation value subtracted $O_{\mathrm{sub}}(\vec{q},\tau) =  [ O(\vec{q},\tau) - \langle  O(\vec{q},\tau)\rangle ]$.

\section{Analysis and Results}
\label{sec:results}
\subsection{Renormalization}

On the lattice, it is shown that the trace anomaly emerges with the lattice regulation after renormalization \cite{CARACCIOLO_1990119,Suzuki_10.1093/ptep/ptu070,DallaBrida:2020gux}. The usual renormalization procedure entails renormalizing the $F^2$ operator on the lattice with the RI/MOM scheme, mixing with other dimension-four and dimension-three operators~\cite{CARACCIOLO_1990119} and matching it to the $\overline{MS}$ scheme at a certain scale. The $\beta/2g$ and $\gamma_m$ factors would then be calculated on this lattice.

Instead of performing this renormalization procedure, we shall take advantage of the mass sum rule and
combine the lattice $\beta/2g$ and the renormalization constant of $F^2$ as an effective $\beta/2g$ factor for the trace anomaly. Since the effective $\beta/2g$ and $\gamma_m$ are independent of the quark mass and the hadron state for the lattice we work on, we shall fit them to several quark masses and hadron states as has been
done in Ref. \cite{Fangcheng_PhysRevD.104.074507}. The only complication in the lattice calculation is that the trace anomaly operators mentioned above can mix with the lower-dimensional operator $\bar{\psi}\psi$ with a $1/a$ power divergence. Several quark masses are needed to fit the $1/a$ term to see if the term is important. In Ref.~\cite{Fangcheng_PhysRevD.104.074507} on the same lattice ensemble, the effective $\beta/2g$ and $\gamma_m$ are calculated with the pseudoscalar and vector meson mass relations based on the sum rule for a valence quark mass. Note that the $1/a$ term has been implicitly included in the sum rule equations. The fact that the sum rule is also satisfied within errors at several valence quark masses indicates that the signal from the $1/a$ term is too weak to be isolated in this calculation \cite{Fangcheng_PhysRevD.104.074507}. 

In this work, on the same lattice ensemble used in Ref.~\cite{Fangcheng_PhysRevD.104.074507}, we only calculate glue contributions. However, we can still obtain $\frac{\beta}{2g}$ from the nucleon mass at one valence mass and it can be used at other valence masses on the same lattice.  Based on the fact that the sigma term in the nucleon is small, we can ignore the $\gamma_m$-related term in the trace anomaly
\begin{align}
\begin{aligned}
    M_{\mathrm{N}}|_{m_\mathrm{v}a} &= (1 + \gamma_m )\langle H_{m} \rangle_{\mathrm{N}} + \frac{\beta(g)}{2g}\langle F^2 \rangle_{\mathrm{N}}|_{m_\mathrm{v}a} \\
    & \approx \langle H_{m} \rangle_{\mathrm{N}} + \frac{\beta(g)}{2g}\langle F^2 \rangle_{\mathrm{N}}|_{m_\mathrm{v}a}.
\end{aligned}
\end{align}
By using the RG invariant sigma term $\langle H_{m} \rangle_{\mathrm{N}}$ which was previously calculated~\cite{Yi-Bo_sigma_piN_PhysRevD.94.054503}, we can obtain the bare $\frac{\beta}{2g}$ on the 24I ensemble
\begin{align}
\begin{aligned}
    \frac{\beta(g)}{2g}|_{m_\mathrm{v}a}  &=  \left. \frac{M_{\mathrm{N}} -  \langle H_{m} \rangle_{\mathrm{N}}}{\langle F^2 \rangle_{\mathrm{N}}}\right \vert_{m_\mathrm{v}a}.
\end{aligned}
\end{align}

Using the data at $m_\mathrm{v}a = 0.016$, we obtained the bare value of $\frac{\beta(g)}{2g}$ to be $-0.129(6)$ and we use this value in all our plots of the glue part of the trace anomaly form factors. The value here is more negative than $-0.08(1)$ obtained in Ref.~\cite{Fangcheng_PhysRevD.104.074507} which uses the pseudoscalar and vector meson masses with $m_q\sim 500$ MeV to determine both $\frac{\beta(g)}{2g}$ and $\gamma_m$. The tension here could be an ${\cal O}(m_q^2a^2)$ discretization error and further investigation on the systematic uncertainty of $\frac{\beta(g)}{2g}$ will be conducted in the future.

\subsection{Three-point function fit}

By including the first excited states, the functional form we used for fitting 3pt is
\begin{align}
\label{eqn:3pt_func_form}
    \begin{aligned}
        C_{\mathrm{H},\mathrm{3pt}}(\vec{p}_i, \vec{p}_f, t,\tau) = &m_{\mathrm{H}}\, G_{\mathrm{H}} (Q^2)\times\\ &\, \mathcal{K}_{\mathrm{H},\mathrm{3pt}}(p_i,p_f)Z_{\vec{p}_i} Z_{\vec{p}_f} e^{-E_i \tau - E_f(t-\tau)} \\
        &+ C_1 e^{-E^1_i \tau - E_f(t-\tau)} \\
        &+ C_2 e^{-E_i \tau - E^1_f(t-\tau)} \\
        &+ C_3 e^{-E^1_i \tau - E^1_f(t-\tau)},
    \end{aligned}
\end{align} 
where $t$ is the source-sink time separation and $\tau$ is the current-source time separation.  $\mathcal{K}_{\mathrm{H},\mathrm{3pt}}(p_i,p_f)$ is the kinematic factor which is listed in Table~\ref{tab:kinematics_partial}.

\begin{table}[!htbp]
    \centering
    \begin{tabular}{|c|c|c|} \hline
        $\mathcal{K}_{\mathrm{H}}$ & $\pi$ & $\mathrm{N}$ \\ \hline
        2pt & $m_{\pi}/E_{\pi,\vec{p}}$ & $(m_N + E_{\mathrm{N},\vec{p}})/E_{\mathrm{N},\vec{p}}$ \\ \hline
        3pt & $m_{\pi}^2/(E_{\pi,\vec{p}_i}E_{\pi,\vec{p}_f})$  & $(m_N + E_{\mathrm{N},\vec{p}_f})/E_{\mathrm{N},\vec{p}_f} $\\ \hline
    \end{tabular}
    \caption{The kinematic factors of the two- and three-point functions for the pion and nucleon used in this work. Note the 3pt kinematic factor for the nucleon is only valid for the case where the source is at rest.}
    \label{tab:kinematics_partial}
\end{table}

We fit the 2pt with zero momentum and use the dispersion relation to calculate the energies. 
$Z_{\vec{p}_i}$ and $Z_{\vec{p}_f}$ are overlap factors between the hadron states and the interpolating operators. The $E$ and $E^1$ are the
ground state and first excited state energies, respectively. In the 3pt$-$2pt joint fit, the parameters $Z_{\vec{p}_i}$$Z_{\vec{p}_f}$ $E_i$ , $E_f$ , $E_{i}^1$, and $E_{f}^1$ are constrained by the joint fit
with the corresponding 2pt.

The associated 2pt is fitted with the functional form
\begin{align}
\label{eqn:2pt_func_form}
    \begin{aligned}
     C_{\pi,\mathrm{2pt}}(\vec{p}_i, \vec{p}_f, 
 t, \tau) = &  \mathcal{K}_{\pi,\mathrm{2pt}}(E) Z_{\vec{p}_i} Z_{\vec{p}_f}  [e^{-E t} + e^{-E(T-t)}]\\
        &  + A_1 e^{-E^1t},\\
     C_{\mathrm{N},\mathrm{2pt}}(\vec{p}_i, \vec{p}_f, 
 t, \tau) = & \mathcal{K}_{\mathrm{N},\mathrm{2pt}}(E) Z_{\vec{p}_i} Z_{\vec{p}_f}  e^{-E t} + A_1 e^{-E^1t},
    \end{aligned}
\end{align}
where $T$ is the lattice size of the time dimension and $A_1$ is a free fitting parameter for the excited-state contributions. 

In addition to the 3pt$-$2pt joint fit method, we can also obtain the glue trace anomaly form factor by calculating the ratio of the 3pt and 2pt functions,
\begin{align}
    \begin{aligned}
        R_{\mathrm{sqrt,H}}
        &(t,\tau;\vec{p}_i,\vec{p}_f) = \frac{ C_{\mathrm{H},\mathrm{3pt}}(t,\tau;\vec{p}_i,\vec{p}_f)}{C_{\mathrm{H},\mathrm{2pt}}(t;\vec{p}_f)} \times  \\ &\sqrt{\frac{C_{\mathrm{H},\mathrm{2pt}}(t-\tau;\vec{p}_i)C_{\mathrm{H},\mathrm{2pt}}(t;\vec{p}_f)C_{\mathrm{H},\mathrm{2pt}}(\tau;\vec{p}_f)}{C_{\mathrm{H},\mathrm{2pt}}(t-\tau;\vec{p}_f)C_{\mathrm{H},\mathrm{2pt}}(t;\vec{p}_i)C_{\mathrm{H},\mathrm{2pt}}(\tau;\vec{p}_i)}} \Bigg/ \\
        &\Bigg[  \frac{m_{\mathrm{H}}\mathcal{K}_{\mathrm{H},\mathrm{3pt}}(p_i,p_f)}{\sqrt{\mathcal{K}_{\mathrm{H},\mathrm{2pt}}(p_i)\mathcal{K}_{\mathrm{H},\mathrm{2pt}}(p_f)} }\Bigg], 
    \end{aligned}
\end{align}
and performing fitting with multiple source-sink separation $t$ with the following functional form of the ratio $R_{\mathrm{sqrt,H}}$,
\begin{align}
\label{eqn:ratio_sqrt_func_form}
    \begin{aligned}
        R_{\mathrm{sqrt,H}}
        & \sim G_{\mathrm{H}}(Q^2) + C''_1 e^{-\Delta E^1_i \tau} +  C''_2 e^{-\Delta E^1_f(t-\tau)} \\
        &\quad \quad \quad+ C''_3 e^{-\Delta  E^1_i \tau - \Delta E^1_f(t-\tau)} \\
        &\xrightarrow[]{t \gg \tau \gg 0} G_{\mathrm{H}}(Q^2),
    \end{aligned}
\end{align}
where the terms with $C_1''$, $C_2''$, and $C_3''$ are the contributions from the excited-state contamination and $\Delta E_{\alpha} = E^1(\vec{p}_{\alpha}) - E(\vec{p}_{\alpha}) $ is the energy difference between the nucleon or pion energy $E(\vec{p}_{\alpha})$ and that of the first excited state $E^1(\vec{p}_{\alpha})$ for either the source (${\alpha}=i$) or the sink (${\alpha}=f$). 
These two methods should yield consistent results. We fit our results with both methods and they are consistent with each other and our fits can describe the data well with $\chi^2/\mathrm{d.o.f.} \sim 1$.

\begin{figure}[!htbp]
    \centering
        \includegraphics[width=0.4\textwidth]{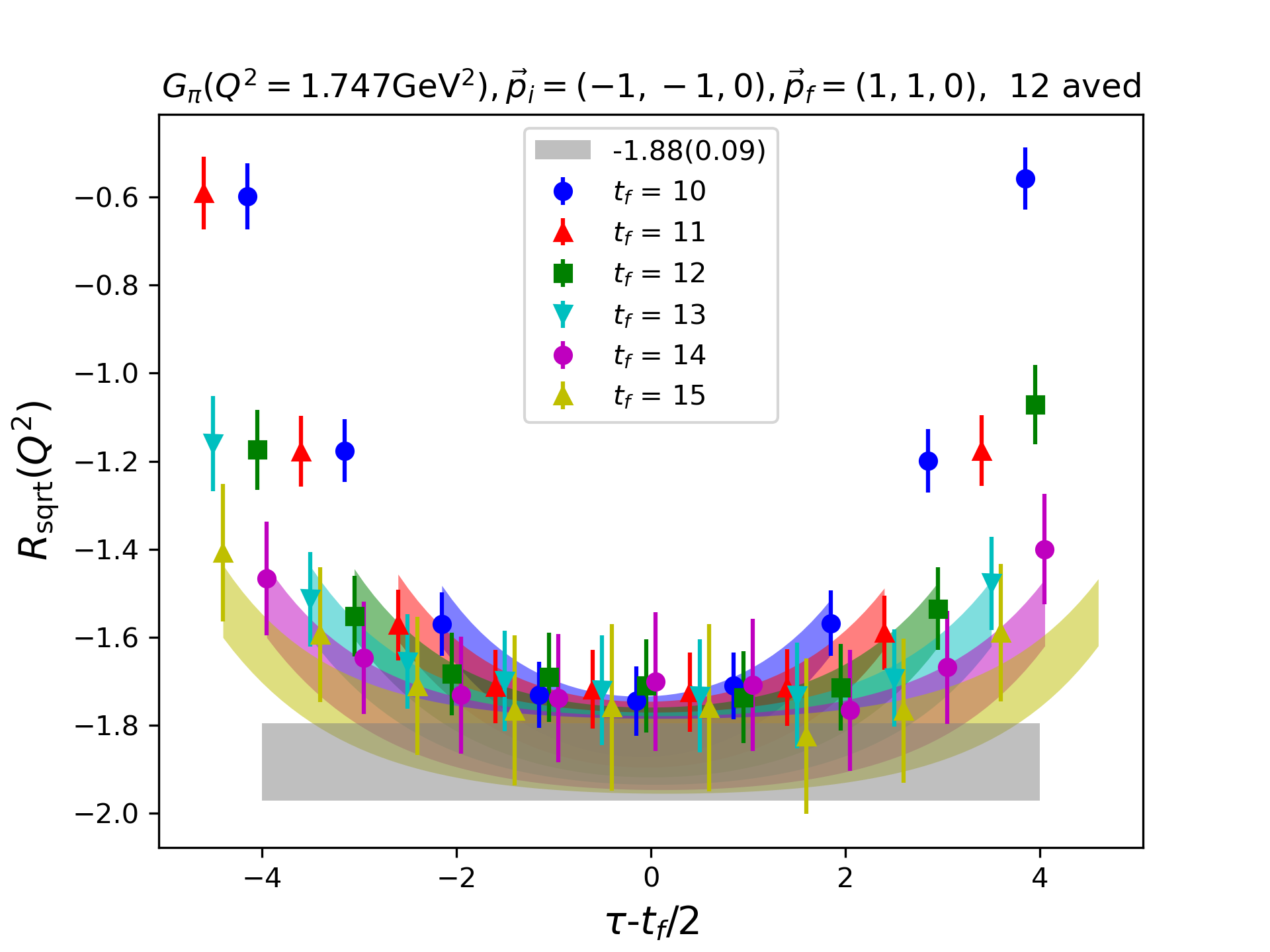}
  \caption{Examples of the ratios of the pion for the special case where $|\vec{p}_i| = |\vec{p}_f|$ on 24I with various values of source-sink separation $t_f$ and current position $\tau$ at the valence pion mass $m_{\pi,\mathrm{v}}$ = 340 MeV. The data points agree well with the colored bands predicted from the fit, and the gray band is for the ground state form factor $G_{\pi}(Q^2)$.} 
\label{fig:3pt_2pt_joint_fit_back_to_back_example}
\end{figure}

\begin{figure}[!htbp]
    \centering
        \includegraphics[width=0.4\textwidth]{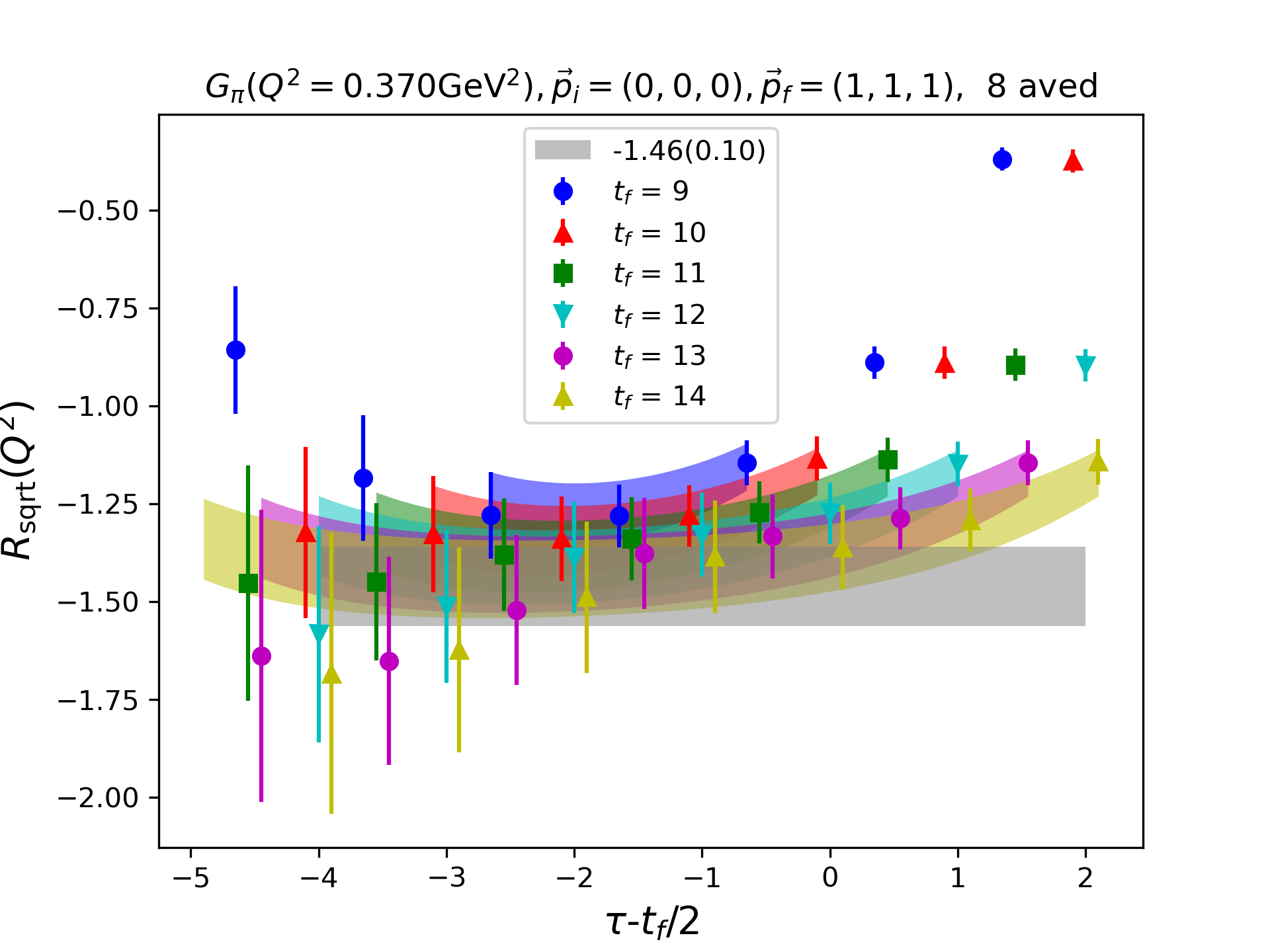}
  \caption{Examples of the ratios of the pion on 24I with various values of source-sink separation $t_f$ and current position $\tau$ at the valence pion mass $m_{\pi,\mathrm{v}}$ = 340 MeV. The plots show the $|\vec{p}_i| \neq |\vec{p}_f|$ case with $\vec{p}_i = 0$, $\vec{q} = \vec{p}_f$. The data points agree well with the colored bands predicted from the fit, and the gray band is for the ground state form factor $G_{\pi}(Q^2)$.} 
\label{fig:3pt_2pt_joint_fit_static_source_example}
\end{figure}

\begin{figure}[!htbp]
    \centering
    \subfloat{\includegraphics[width=0.4\textwidth]{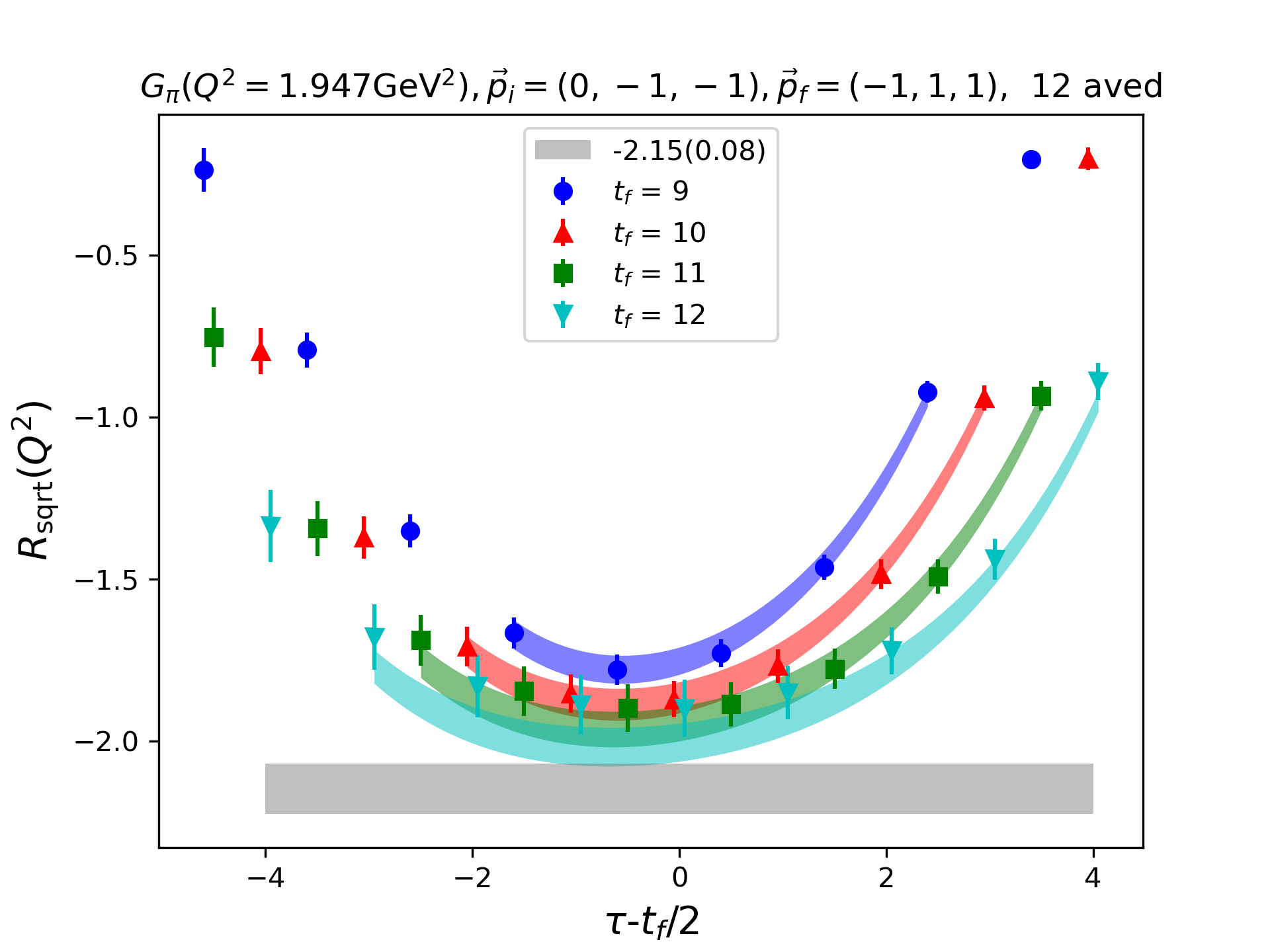} }
    \label{fig:near_back_to_back_i7Q2_1.947}
    \vspace{-10pt}
     
     \subfloat{\includegraphics[width=0.4\textwidth]{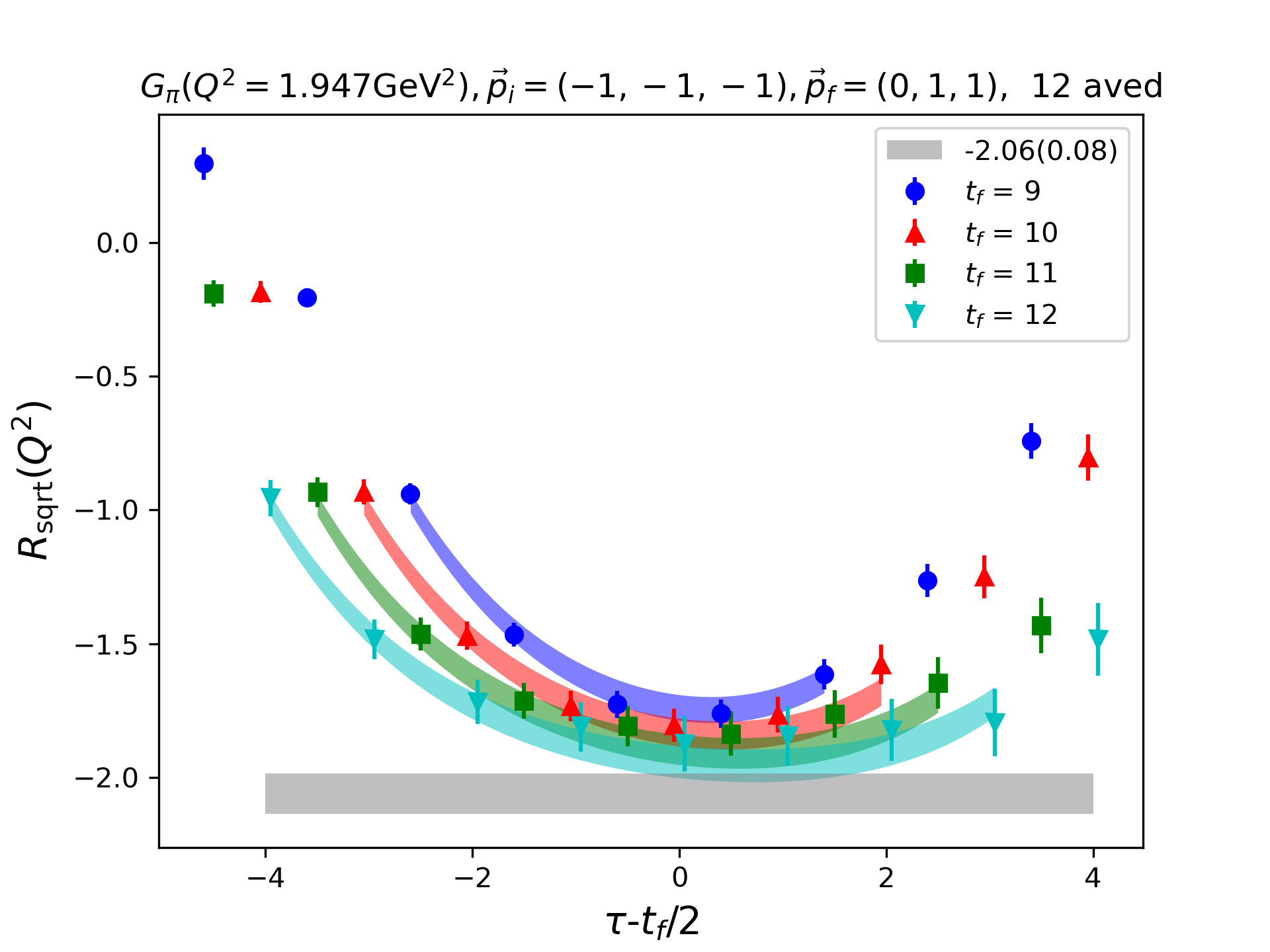} }
    \label{fig:near_back_to_back_i6Q2_1.947}

  \caption{Examples of the ratios of the pion on 24I with various values of source-sink separation $t_f$ and current position $\tau$ at the valence pion mass $m_{\pi,\mathrm{v}}$ = 340 MeV. The plots show the general $|\vec{p}_i| \neq |\vec{p}_f|$ case with near-back-to-back momentum transfer. The data points agree well with the colored bands predicted from the fit, and the gray band is for the ground state form factor $G_{\pi}(Q^2)$.} 
\label{fig:3pt_2pt_joint_fit_near_back_to_back_example}
\end{figure}

We show the pion results with three types of momentum transfers in Fig.~\ref{fig:3pt_2pt_joint_fit_back_to_back_example} (back to back), Fig.~\ref{fig:3pt_2pt_joint_fit_static_source_example} (source at rest), and Fig.~\ref{fig:3pt_2pt_joint_fit_near_back_to_back_example} (near back to back). For illustrative purposes, the data points on the panels are shown with ratio $R_{\mathrm{sqrt,H}}$ and the reconstructed ratios based on the functional form described in Eqs. (\ref{eqn:3pt_func_form}), (\ref{eqn:2pt_func_form}), and (\ref{eqn:ratio_sqrt_func_form}) are shown with colored bands. The data points are fitted well ($\chi^2/\mathrm{d.o.f.} \sim 1$) and the reconstructed ratios are consistent with the data. For the back-to-back case, we find that the data points are symmetric about $\tau = t/2$ within uncertainty. In addition, for the near-back-to-back case, the data points are symmetrical within uncertainty upon the exchange of source-sink momentum. We can therefore confirm that the source smearing with momentum applied by phase factors for grid sources and sink smearing with momentum have the same overlap with the pion states.

\begin{figure}[!htbp]
    \centering
        \includegraphics[width=0.4\textwidth]{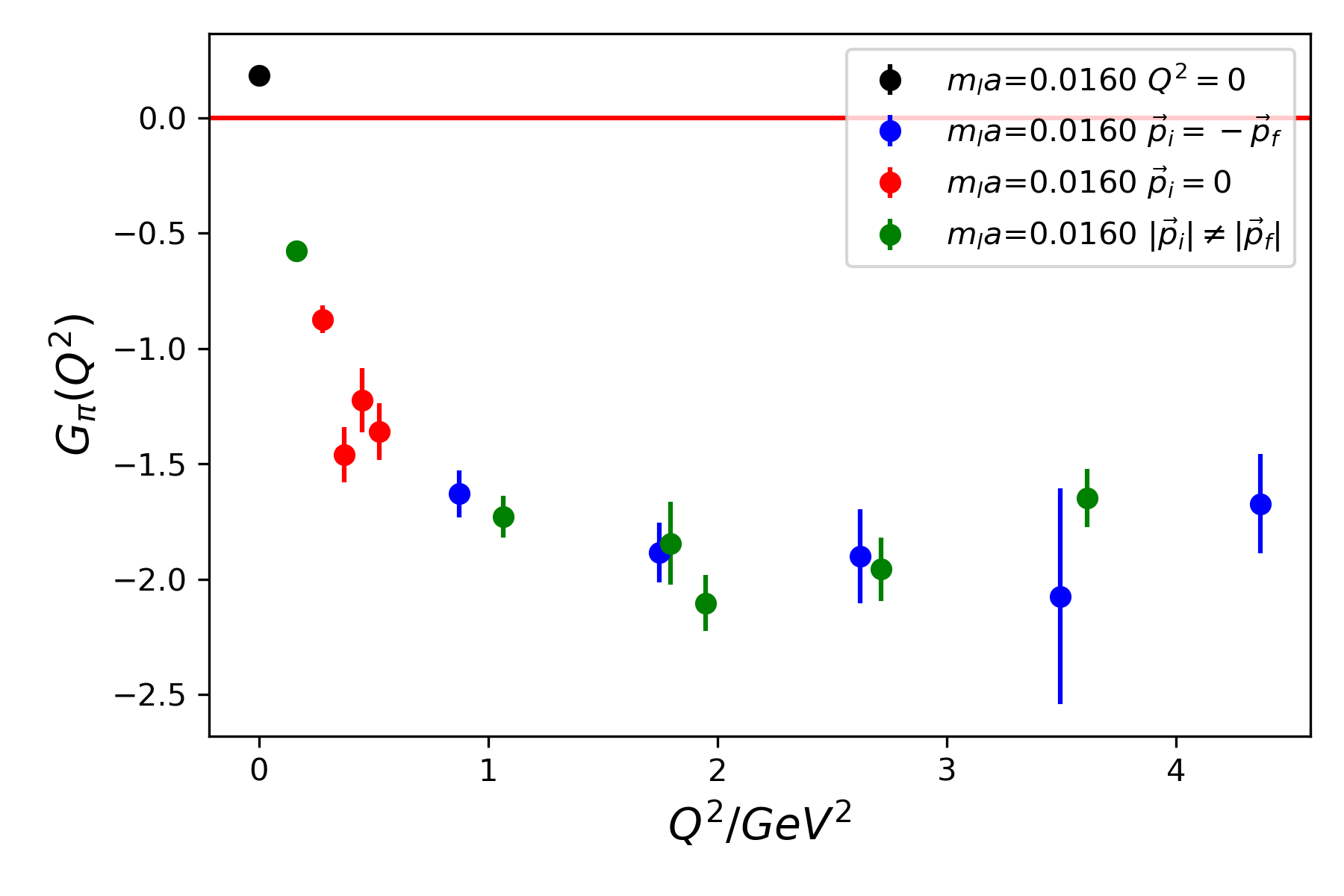}
        \caption{The glue trace anomaly form factor $G_{\pi}(Q^2)$ with $m_la = 0.016$, which corresponds to $m_{\pi, \mathrm{v}}=~340$ MeV. Different momentum transfer scenarios are plotted with different colors.} 
        \label{Fig:pion_trace_anomaly_FF}
\end{figure}

\begin{figure}[!htbp]
    \centering
        \includegraphics[width=0.4\textwidth]{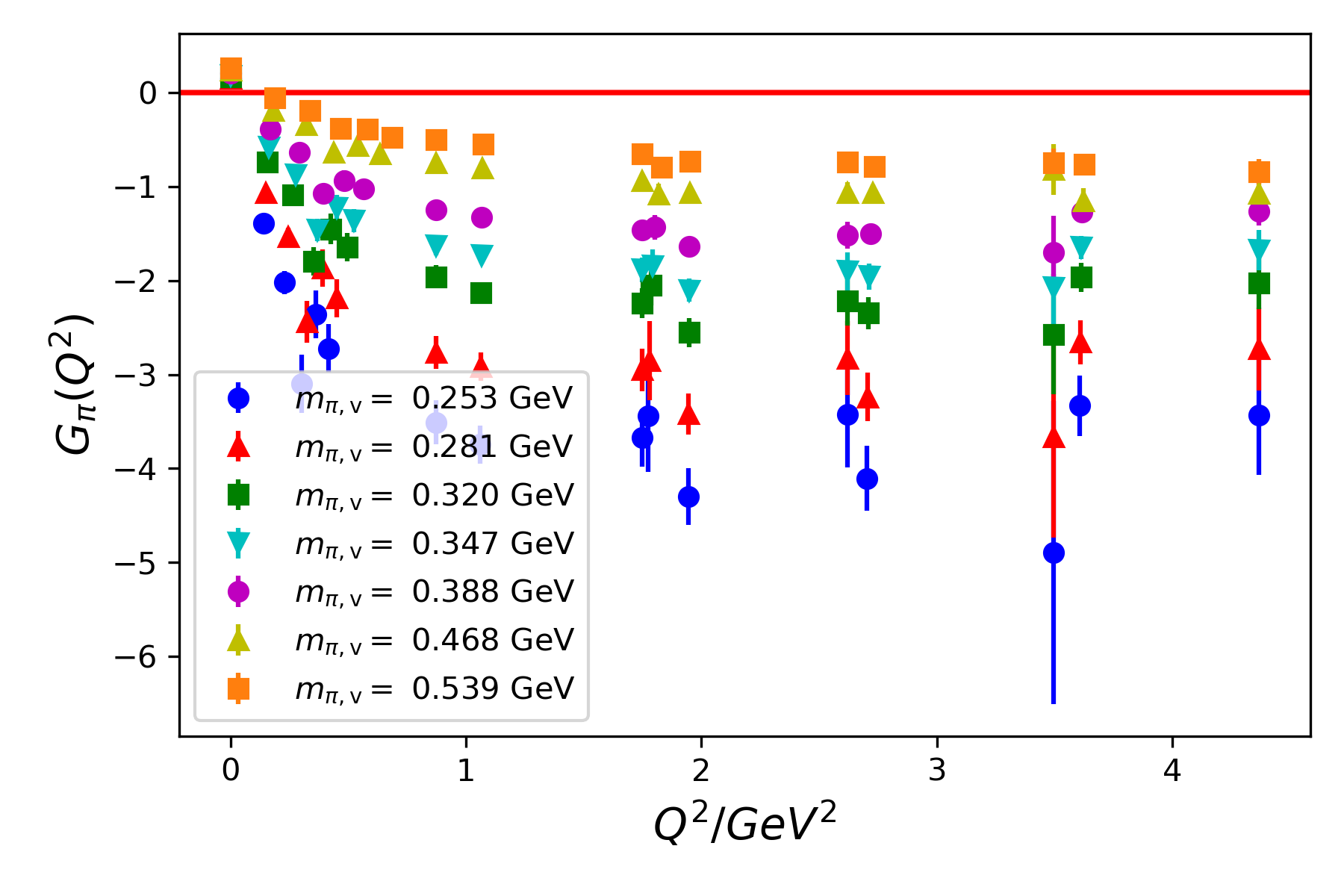}
        \caption{The glue trace anomaly form factor $G_{\pi}(Q^2)$ at seven valence pion masses. } \label{Fig:pion_trace_anomaly_FF_multimass}
\end{figure}

\begin{figure}[!htbp]
    \centering
        \includegraphics[width=0.4\textwidth]{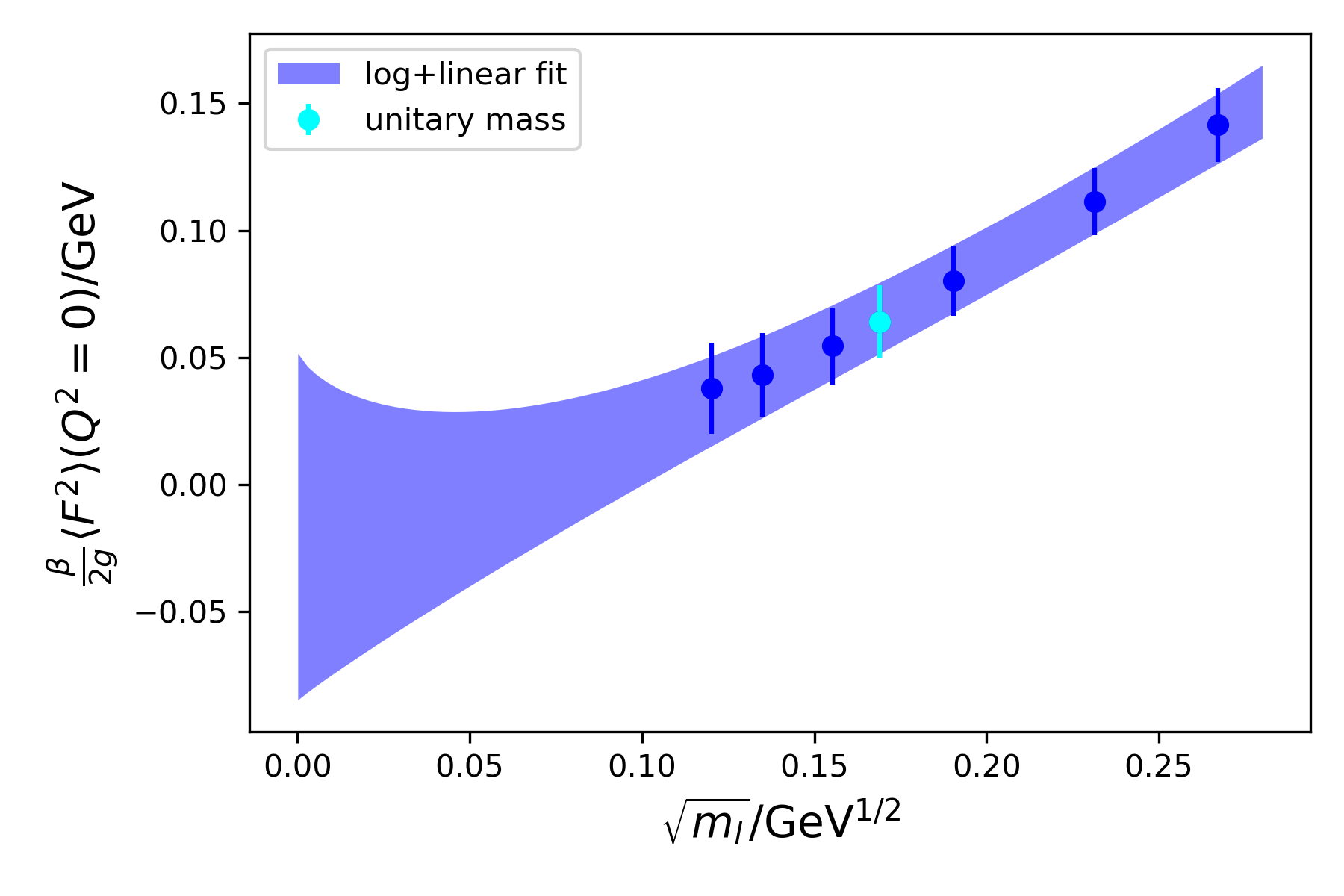}
        \caption{The glue trace anomaly matrix element of the pion, $\frac{\beta}{2g}\langle F^2 \rangle (Q^2 = 0)$, in the forward limit (i.e., at $Q^2=0~\mathrm{GeV}^2$). The data points agree well with the colored bands predicted from the fit with the functional form 
        $\frac{\beta}{2g}\langle F^2 \rangle=a+b\sqrt{m_l}+c\sqrt{m_l}\ln(\sqrt{m_l})$.}
        \label{Fig:pion_F2_multimass}
\end{figure}

By combining the data from all the momentum transfer scenarios, we obtain the final results for the form factors. For the pion, we plot $G_{\pi}(Q^2)$ for the unitary quark mass in Fig.~\ref{Fig:pion_trace_anomaly_FF} and plot the results at all seven valence quark masses in Fig.~\ref{Fig:pion_trace_anomaly_FF_multimass}. Here are several important features we find from the results. First, in the forward limit (i.e., at $Q^2=0~\mathrm{GeV}^2$), we get a positive value, which is consistent with the glue contribution to the hadron mass~\cite{Fangcheng_PhysRevD.104.074507}. We plot the forward matrix element $\frac{\beta}{2g}\langle F^2 \rangle (Q^2 = 0)$ with respect to $\sqrt{m_l}$ in Fig.~\ref{Fig:pion_F2_multimass} and find that the glue part of the pion trace anomaly is proportional to $\sqrt{m}$ for quark mass above and at the unitary mass, just as are the pion mass and its sigma term~\cite{liu2023hadrons}. Second, as the $Q^2$ gets larger, the form factor becomes negative and there is a sign change which is predicted~\cite{liu2023hadrons} based on the sign change observed in the spatial distribution of the trace anomaly~\cite{Fangcheng_PhysRevD.104.074507}. 

At small $Q^2$, chiral perturbation theory (ChPT) is useful in terms of predicting the trace anomaly form factors. The derivation in Ref.~\cite{Novikov:1980fa} using the soft-pion theorem, the chiral perturbation calculation of the  trace anomaly matrix element at tree level~\cite{Chen_ChPT_TAFF_1998}, and the calculation of the trace anomaly form  factor of pion using the gravitational form factors~\cite{Hatta_note_Chiral_Perturb_Soft_Pion_Small_Q2}, all yield
\begin{equation}
    \label{eqn:TAFF_expression_ChPT}
    \mathcal{F}_{\mathrm{ta},\pi}^{\mathrm{ChPT}}(Q^2) \sim \frac{1}{2} - \frac{1}{2m_{\pi}^2}Q^2.
\end{equation}

On the one hand, from Eq.~(\ref{eqn:TAFF_expression_ChPT}), the form factor is positive at $Q^2 = 0$ with $\mathcal{F}_{\mathrm{ta},\pi}^{\mathrm{ChPT}}(Q^2 = 0)=\frac{1}{2}$ and on the other hand, Eq.~(\ref{eqn:TAFF_expression_ChPT}) predicts a sign change in the trace anomaly form factor of the pion. Therefore our results are consistent with the predictions from the chiral perturbation theory at the small $Q^2$ region.

\begin{figure}[!htbp]
    \centering
        \includegraphics[width=0.4\textwidth]{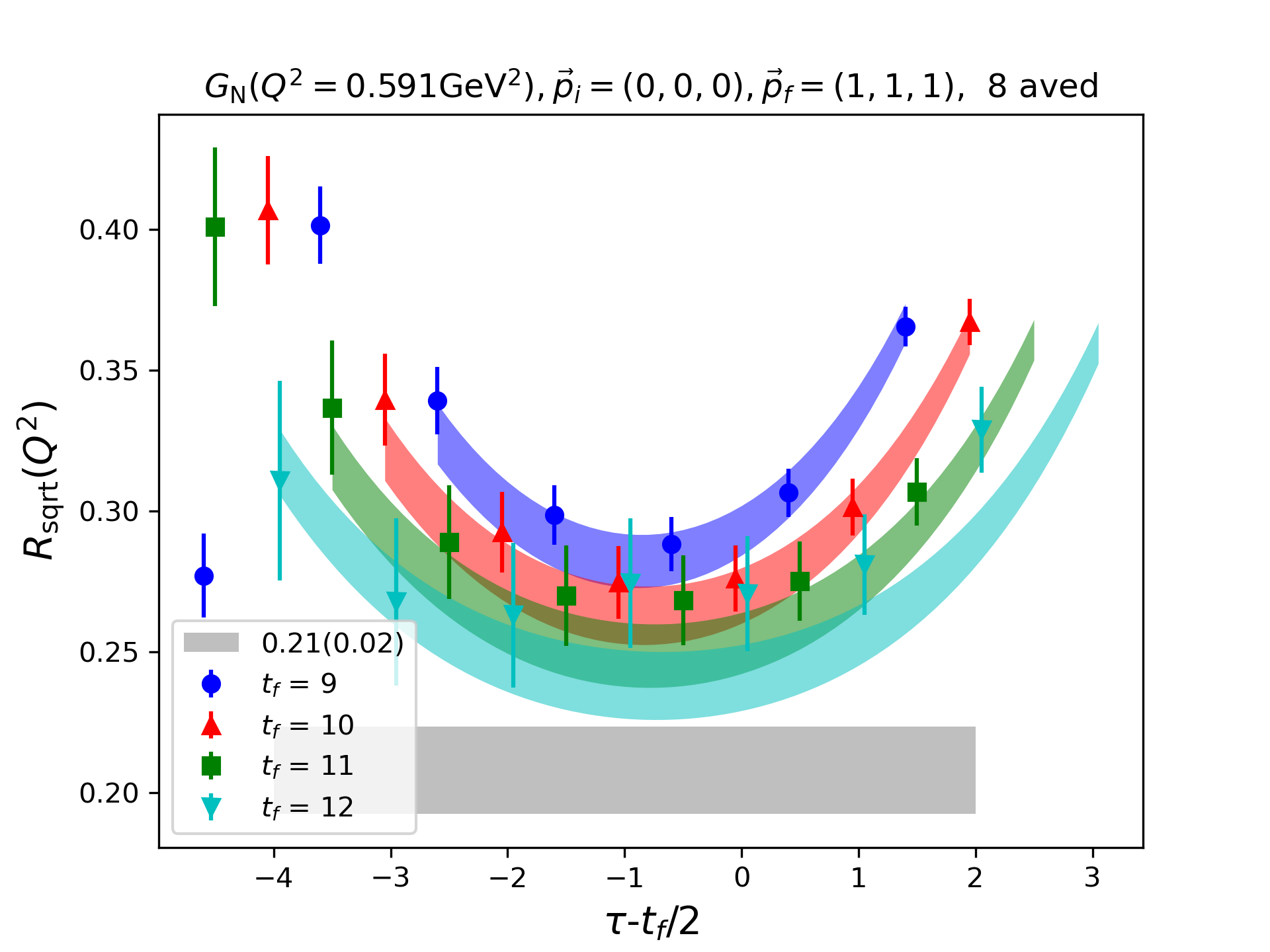}
  \caption{Examples of the ratios of the nucleon on 24I with various values of source-sink separation $t_f$ and current position $\tau$ at the valence pion mass $m_{\pi,\mathrm{v}}$ = 340 MeV. The plots show the $|\vec{p}_i| \neq |\vec{p}_f|$ case with $\vec{p}_i = 0$, $\vec{q} = \vec{p}_f$. The data points agree well with the colored bands predicted from the fit, and the gray band is for the ground state form factor $G_{N}(Q^2)$. } 
\label{fig:3pt_2pt_joint_fit_proton_static_source_example}
\end{figure}

\begin{figure}[!htbp]
    \centering
        \includegraphics[width=0.4\textwidth]{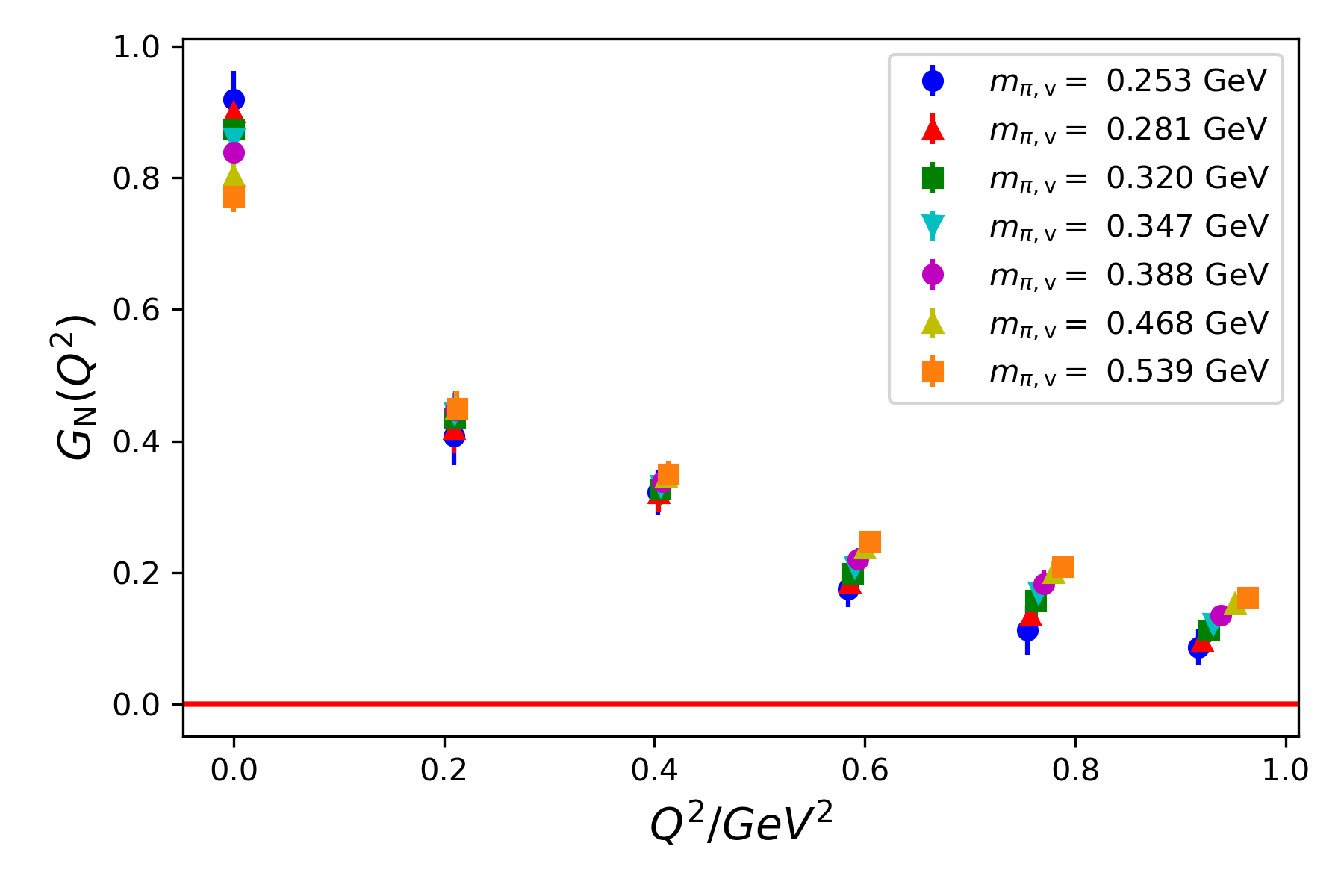}
        \caption{The glue trace anomaly form factor $G_{N}(Q^2)$ at seven valence pion masses. } 
        \label{Fig:proton_trace_anomaly_FF_multimass}
\end{figure}

We also calculate $G_N(Q^2)$ for the nucleon. We show the nucleon results with the source at rest in Fig.~\ref{fig:3pt_2pt_joint_fit_proton_static_source_example} and the results for the seven valence pion masses are plotted in Fig.~\ref{Fig:proton_trace_anomaly_FF_multimass}. In contrast to the pion case, $G_N(Q^2)$ decreases monotonically as $Q^2$ increases and there is no sign change. This is consistent with the expectation from the trace anomaly density of the nucleon where no sign change is found~\cite{Fangcheng_PhysRevD.104.074507}. A recent perturbative QCD calculation of the trace anomaly form factors at large $Q^2$ predicted the asymptotic signs for the pion and nucleon ~\cite{Tong:2022zax}. Their results agree with this work for the case of the pion but disagree with this work for the case of the nucleon.

\subsection{$z$-expansion fit of pion and nucleon form factors}
In order to obtain the trace anomaly radius and spatial distribution, we perform a model-independent $z$-expansion fit~\cite{z-expansion_PhysRevD.82.113005} with the ratio $\tilde{G}_{\mathrm{H}}(Q^2) = G_{\mathrm{H}}(Q^2)/G_{\mathrm{H}}(Q^2=0)$ for pion and nucleon separately using the following equation
\begin{equation}
    \tilde{G}_{\mathrm{H}}(Q^2) = \tilde{\mathcal{G}}_{\mathrm{H}}(z) = \sum^{k_{\mathrm{max}}}_{k=0}a_k z^k,
\end{equation}
\begin{equation}
    z(t,t_{\mathrm{cut}},t_0) = \frac{\sqrt{t_{\mathrm{cut}}-t}-\sqrt{t_{\mathrm{cut}}-t_0}}{\sqrt{t_{\mathrm{cut}}-t}+\sqrt{t_{\mathrm{cut}}-t_0}},
    \label{eqn:map_t_2_z}
\end{equation}
where $t=-Q^2$, and $t_{\mathrm{cut}}$ is set to be at the two-pion threshold, i.e.,  $t_{\mathrm{cut}}=4m_{\pi,\mathrm{mix}}^2$, with $m_{\pi,\mathrm{mix}}$ being the mass of the mixed valence and sea pseudoscalar meson on this ensemble as calculated in Ref.~\cite{Delta_mix_PhysRevD.86.014501}; and $t_0$ is chosen to be its ``optimal" value $t^{\mathrm{opt}}_0(Q_{\mathrm{max}})=t_{\mathrm{cut}}(1-\sqrt{1+Q_{\mathrm{max}}^2/t_{\mathrm{cut}}})$ to minimize the maximum value of $|z|$, with $Q_{\mathrm{max}}^2$ the maximum $Q^2$ under consideration.

The model dependence of the fit can be significantly suppressed by reaching to higher $k_{\mathrm{max}}$ where the fitting results are insensitive to the values of the choice of $k_{\mathrm{max}}$. In order to achieve a higher $k_{\mathrm{max}}$, we add constraints to the parameters based on the asymptotic behaviors of the form factors in the limit $Q^2 \rightarrow \infty$.

Using Eq.~(\ref{eqn:map_t_2_z}), in the limit $Q^2 \rightarrow \infty$, we have $z \rightarrow 1$ and $1-z = \frac{1}{Q}$ and the form factor
\begin{align}
    \begin{aligned}
        \tilde{\mathcal{G}}(z) & = \tilde{\mathcal{G}}(1) + \sum_{n} \left.\frac{d^n\tilde{\mathcal{G}}}{dz^n}\right\vert_{z=1}(1-z)^n \\
        &\sim \tilde{\mathcal{G}}(1) + \sum_{n} \left.\frac{d^n\tilde{\mathcal{G}}}{dz^n}\right\vert_{z=1} \bigg (\frac{1}{Q^n} \bigg ).
    \end{aligned}
    \label{eqn:asymp_expand_z}
\end{align}
If we assume the form factors fall as powers of $\frac{1}{Q}$ at large $Q^2$, then we can write $\tilde{\mathcal{G}}(z\rightarrow 1) \propto 1/Q^l$, where $l$ is an integer. Combining this with Eq.~(\ref{eqn:asymp_expand_z}), we obtain the constraints for the coefficients $a_k$
\begin{equation}
    \tilde{\mathcal{G}}(1) =\sum^{\infty}_{k=0}a_k = 0
    \label{eqn:zxep_power_constrain_0},
\end{equation}
\begin{equation}
    \left. \frac{d^n\tilde{\mathcal{G}}}{dz^n}\right\vert_{z=1}= \sum^{\infty}_{k=n}\frac{k!}{(k-n)!}a_k = 0,~n \in \{1, ..., l-1\}.
    \label{eqn:zxep_power_constrain}
\end{equation}
In this work we take $l=4$ and set $k_{\mathrm{max}}=7$ and $k_{\mathrm{max}}=6$  for the pion and the nucleon respectively, by using \{$a_0$, $a_1$, $a_2$\} as the free parameters and having the rest of the coefficients \{$a_3$, ..., $a_6$/$a_7$\} constrained according to Eq.~(\ref{eqn:zxep_power_constrain_0}) and (\ref{eqn:zxep_power_constrain}).

\begin{figure}[!htbp]
    \centering
        \includegraphics[width=0.4\textwidth]{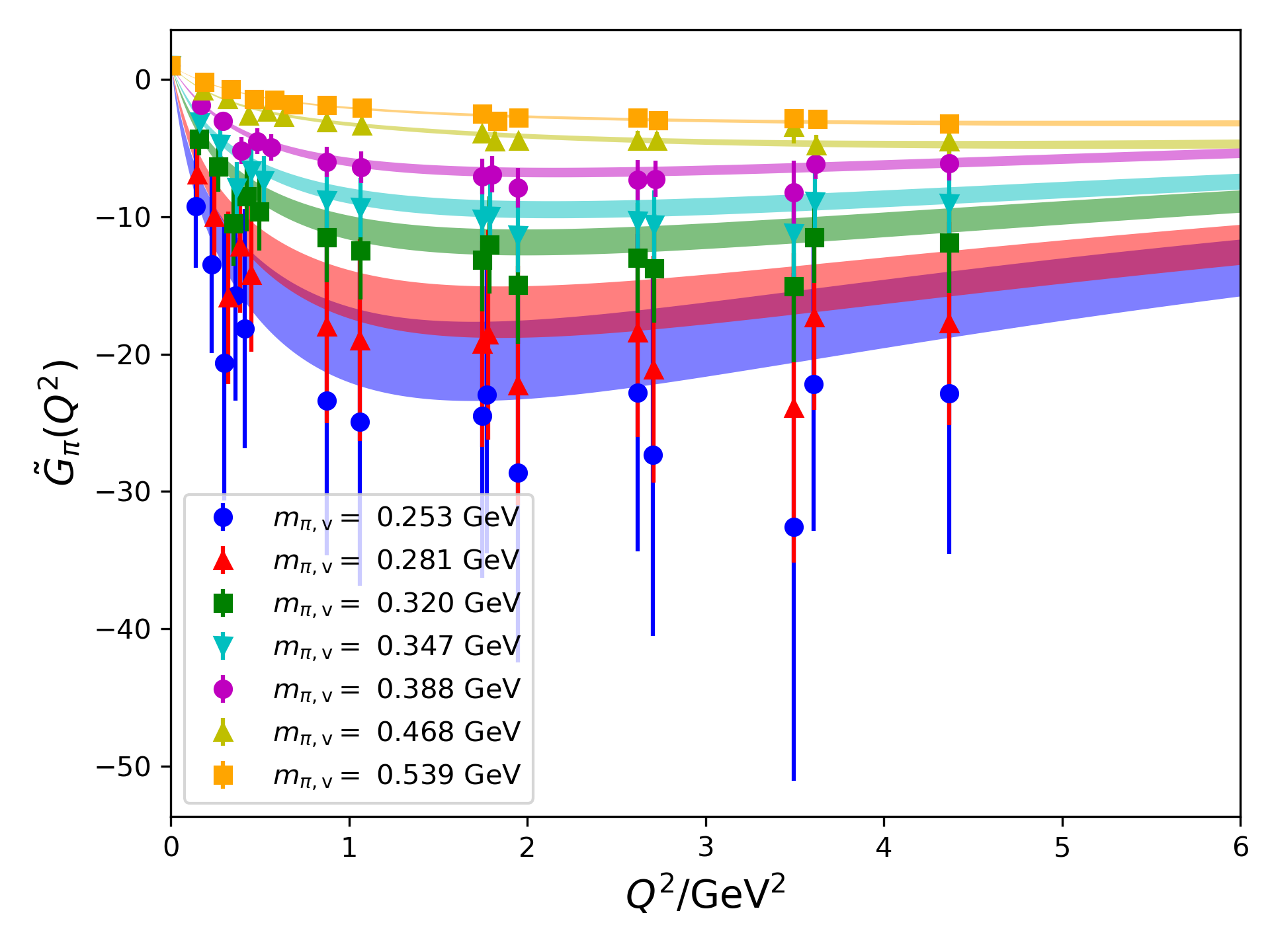}
        \caption{$\tilde{G}_{\pi}(Q^2)$ at seven valence pion masses. The results from $z$-expansion fits are plotted in colored bands.} 
        \label{Fig:pion_trace_anomaly_FF_multimass_normed}
\end{figure}

\begin{figure}[!htbp]
    \centering
        \includegraphics[width=0.4\textwidth]{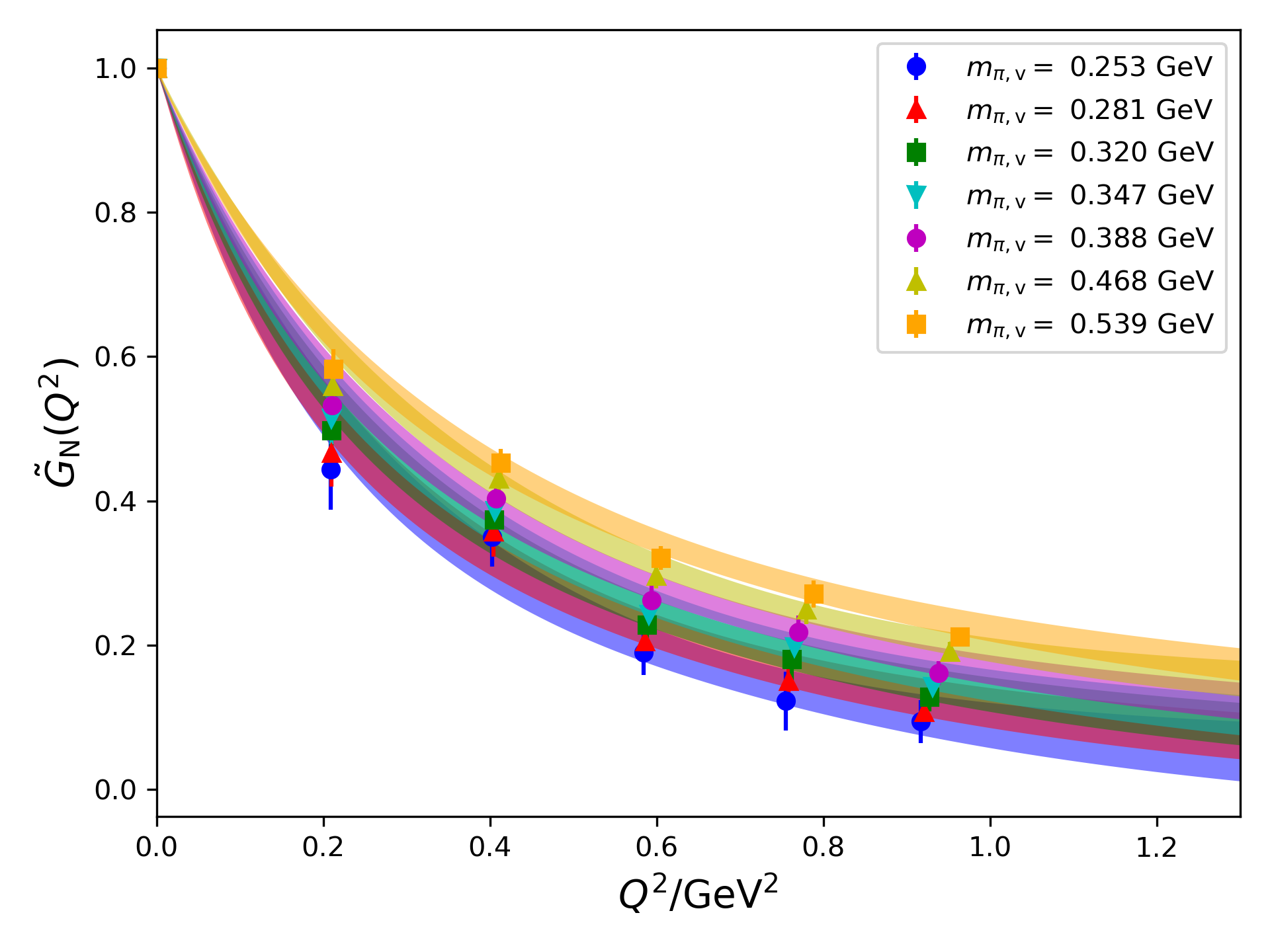}
        \caption{$\tilde{G}_{N}(Q^2)$ at seven valence pion masses. The results from $z$-expansion fits are plotted in colored bands.} 
        \label{Fig:proton_trace_anomaly_FF_multimass_normed}
\end{figure}

The $z$-expansion fitted pion form factors up to $Q^2 \sim4.3~\mathrm{GeV}^2$ for the seven valence quark masses on the 24I ensemble, are shown in Fig.~\ref{Fig:pion_trace_anomaly_FF_multimass_normed} with $\chi^2/\mathrm{d.o.f.} \sim 1$. We also fit the nucleon form factors and the fit results are shown in Fig.~\ref{Fig:proton_trace_anomaly_FF_multimass_normed} with $\chi^2/\mathrm{d.o.f.} \sim 1$. 

\subsection{Chiral extrapolation of the glue trace anomaly radii of the pion and nucleon}
The average squared mass radius of a hadron can be defined as
\begin{equation}
    \label{eqn:mass_radius}
    \langle r^2 \rangle_{\rm m} (\mathrm{H}) = -6 \left. \frac{d\mathcal{F}_{\mathrm{m},\mathrm{H}} (Q^2)}{dQ^2} \right\vert_{Q^2 \rightarrow 0}.
\end{equation}
Since the sigma term is small in the nucleon and its derivative with respect to $Q^2$ is expected to be negligible in the pion as compared to the contribution from the glue part of the trace anomaly, $\langle r^2_g \rangle_{\rm ta} (\mathrm{H}) = -6~dG_{\mathrm{H}}(Q^2)/dQ^2|_{Q^2 \rightarrow 0}$ should be good approximations to the nucleon and pion mass radii and we can use the fitted form factors to obtain $\langle r^2_g \rangle_{\rm ta} (\mathrm{H})$. Based on the ChPT prediction shown in Eq.~(\ref{eqn:TAFF_expression_ChPT}), the pion mass radius is
\begin{equation}\label{eq:radius_chipt}
    \langle r^2\rangle^{\mathrm{ChPT}}_{\rm m} (\pi) = \frac{3}{m_{\pi}^2}.
\end{equation}
 This suggests that the radius of the pion diverges at the chiral limit quadratically in the pion mass and at the physical pion mass,  $\langle r^2 \rangle^{\mathrm{ChPT}}_{\rm m} (\pi) \simeq 6~\mathrm{fm^2}$. This also suggests that the radius of the pion should incorporate a term inversely proportional to $m_\pi^2$ in the chiral extrapolations. Inspired by this observation, we adopt the following ansatz to extrapolate the pion radius to the physical point, 
 \bea
\label{eq:chiral_exp_radius_pion2}
\langle r^2_g \rangle_{\mathrm{ta}}(\pi)&=a_\pi/m_\pi^2+b_\pi +c_\pi\log\left(\frac{m_\pi^2}{m_{\pi,\mathrm{phy}}^2}\right)
\nonumber\\ &+d_\pi m_\pi^2,
\eea
where the chiral-log term is from the prediction of a ChPT calculation~\cite{Chen_ChPT_TAFF_1998}. Our result for the square radius of the pion at the physical point obtained from the fit \textit{Ansatz} is around 21.5(5.2)\,fm$^2$ with $\chi^2/\mathrm{d.o.f.}$ about 1.2. Alternatively, dropping the $d_\pi m_\pi^2$ term results in a radius at the physical point of 9.8(2.1)\,fm$^2$ with $\chi^2/\mathrm{d.o.f.}$ about 2. If we take the difference of results obtained using the different fit \textit{Ansatz} as the systematic error, then our final result is $\langle r^2_g \rangle^{\mathrm{phy}}_{\mathrm{ta}}(\pi)$=21.5(5.2)(11.7)\,fm$^2$.
 In addition to the systematic error from different \textit{Ans\"atze}, please note there are three sources of systematic error which we need to take into account and investigate in the future: (1) the unphysical sea quark mass of the lattice ensemble: a direct calculation at the physical point is necessary; (2) the finite-lattice-spacing effects in the renormalization: the mixing with the lower-dimensional operator and the validity of the sum rule equation upon extrapolation to the continuum limit must be examined; and (3) the finite-volume effects. These systematic errors should also be included in the results for the nucleon in the following passages.

\begin{figure}[!htbp]
    \centering
    \includegraphics[height=5cm,width=0.8\linewidth]{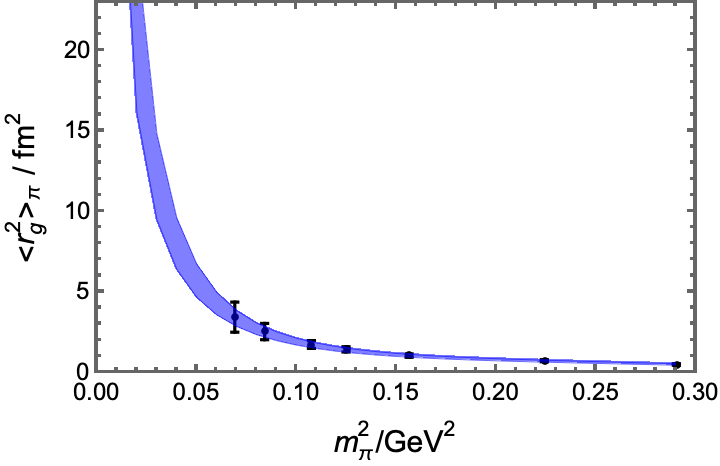} \\
    \includegraphics[height=5cm,width=0.8\linewidth]{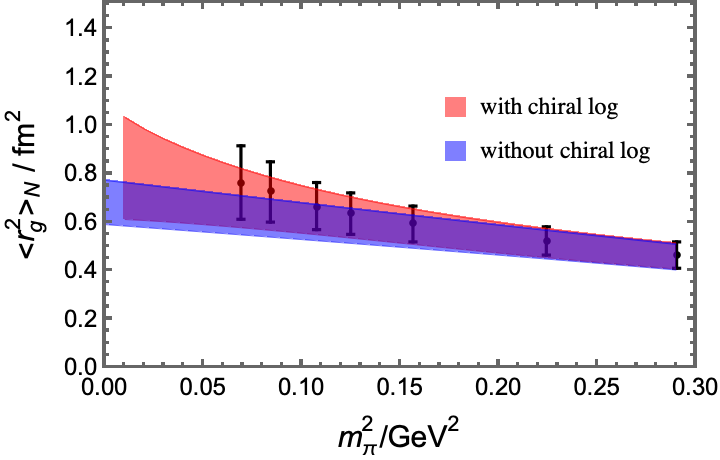}
    \caption{The valence pion mass dependence of the radius of the glue trace anomaly in the pion (top panel) and nucleon (bottom panel). The band in the top panel for the pion case represents the result constructed using Eq.~(\ref{eq:chiral_exp_radius_pion2}). The bands in the bottom panel for the nucleon case are constructed using Eq.~(\ref{eq:chiral_exp_radius}).} 
        \label{Fig:chiral_exp}
\end{figure}

On the other hand, the square radius of the nucleon increases with decreasing pion mass, as inferred from the slope at small $Q^2$ region depicted in  Fig.~\ref{Fig:proton_trace_anomaly_FF_multimass_normed}. To estimate the results at the physical pion mass, we use the following \textit{Ans\"atze} for chiral extrapolation
\bea\label{eq:chiral_exp_radius}
\langle r^2_g \rangle_{\mathrm{ta}}(\mathrm{N})&=&a_{\mathrm{N}}+b_{\mathrm{N}}~m_\pi^2\nonumber\\
\langle r^2_g \rangle_{\mathrm{ta}}(\mathrm{N})&=&a_{\mathrm{N}}+b_{\mathrm{N}}~m_\pi^2+c_{\mathrm{N}}~m_\pi^2 \log\left(\frac{m_\pi^2}{m_{\pi,\mathrm{phy}}^2}\right).
\eea
The pion mass dependence of the radius is shown in the bottom panel of Fig.~\ref{Fig:chiral_exp}. The extrapolated results for the $z$-expansion fit are $\langle r_g^2\rangle^{1/2}_{\rm ta}(N)$ = 0.89(10) $\mathrm{fm}$ and 0.82(05) $\mathrm{fm}$ with and without the chiral log term, respectively. Our final result is 0.89(10)(07) fm if the difference of results obtained using different fit \textit{Ans\"atze} is taken as the systematic error. The radius of the pion being larger than that of the nucleon is consistent with the observation in a recent lattice calculation performed by some of the authors within this study~\cite{Fangcheng_PhysRevD.104.074507}.

The glue trace anomaly form factors can be extracted from the differential cross sections of the $J/\Psi$ photoproduction~\cite{Luke:1992tm,Kharzeev:1995ij,Kharzeev:1998bz,Duran:2022xag} and leptoproduction at large photon virtualities~\cite{Boussarie:2020vmu}. A recent direct dipole fit to the recent GlueX collaboration data~\cite{GlueX:2019mkq} results in a root-mean-square radius of the nucleon of about $0.55(3)$ fm~\cite{Kharzeev:2021qkd}, which is substantially smaller than the present lattice result of $0.89(10)(07)$ fm. Since the nucleon sigma term is neglected, this radius is the
mass radius of the nucleon.~\footnote{The forward matrix element of the trace $T_{\mu}^{\mu}$ and that of $T^{00}$ give 
the same result and yet their radii differ. Since the forward trace matrix element corresponds to the mass of the hadron and that of $T^{00}$ corresponds to the energy, it is appropriate to call the radius from the trace of the GFF the mass radius and that from the GFF of $T^{00}$ the energy radius.}


Another way to access the trace anomaly form factor is through the gravitational form factors (GFFs) which are the moments of GPD~\cite{Ji:2021mtz,liu2023hadrons}. As shown in Eq.~(\ref{G_m}), the EMT trace form factor $\mathcal{F}_{\mathrm{m}}$ is made up
of two parts. It is shown that the EMT trace form factor is related to the GFF~\cite{Ji:2021mtz,liu2023hadrons}, and for the nucleon, one can write
\begin{align} \label{G_ta}
    \begin{aligned}
        \mathcal{F}_{\rm{ta}}(Q^2) =  &\big [( A(Q^2)   - B(Q^2) \frac{Q^2}{4m_{\mathrm{N}}^2 } + 3 D (Q^2)\frac{Q^2}{m_{\mathrm{N}}^2 } \big ]\\ & - \mathcal{F}_{\sigma} (Q^2),
    \end{aligned}
\end{align}
where $A(Q^2), B(Q^2)$, and $D(Q^2)$
are the sum of the quark and glue GFF. This relation is due to the conservation of the EMT, i.e., $\partial_{\mu} T^{\mu\nu} = 0$
~\cite{liu2023hadrons}. The square mass radius of the trace anomaly can then be defined~\cite{Ji:2021mtz} as
\begin{equation} 
\label{r_GFF}
\langle r^2\rangle_{\mathrm{ta}} = - 6\, ( \frac{d A(Q^2)}{dQ^2} +\frac{3 D(0)}{M^2} - \frac{d \mathcal{F}_{\sigma}(Q^2)}{dQ^2}),
\,\,\,\,\,\,\,\,\,\,\,\,\,\,
\end{equation}
where $B(0) = 0$ has been used. For the nucleon, the contribution from the sigma term is negligible and the last term may be dropped in the calculations of the mass radius.
A lattice calculation has been carried out recently to calculate the quark and glue GFF~\cite{Hackett:2023rif} of the nucleon. The mass rms radius from Eq.~(\ref{r_GFF}) is obtained as 1.038(98) fm~\cite{Pefkou23}. This is consistent with our direct calculation of 0.89(10)(07) fm.  The holographic calculation~\cite{Mamo:2022eui} with lattice input
results in $0.926 \pm 0.008~$fm which is also consistent with our result. Using some lattice input for the quark contributions
in Eq.~(\ref{r_GFF}) and the glue GFF from fitting the near-threshold $J/\Psi$ production at $\xi >0$, it is found the scalar radius to be 1.20(13) fm~\cite{Guo:2023pqw} which is slightly larger than the present work. 
Attempts have also been made to extract it from the gravitational form factors $A$ and $D$ from $J/\Psi$ photoproduction near the threshold~\cite{Duran:2022xag}. So far, only the glue part of the $A$ and $D$ are included. As we can see from Eq.~(\ref{G_ta}), both the glue and quark GFF are needed.

\subsection{Spatial distributions of the glue trace anomaly in the pion and nucleon}

After obtaining the fitted form factors, we can calculate the spatial distribution in the instant form of the mass, especially the glue part of the trace anomaly, in a specific frame where the energy transferred to the system $\Delta^0=0$. 

Traditionally, the spatial distributions are defined in the Breit frame where there is no energy transfer, i.e., $\Delta^0 = 0$ and  $\mathbf{P} = (\vec{p} + \vec{p}')/2 = \mathbf{0}$ and therefore $Q^2 = |\mathbf{\Delta}|^2$. For example, the charge densities in the Breit frame can be obtained from the electromagnetic form factors with Fourier transforms. However, it has been pointed out that such a Fourier transform relationship between form factors and spatial distributions of the expectation values of local operators is only valid for nonrelativistic systems and is not accurate for systems whose size is of the order of its Compton wavelength \cite{Burkardt:2000za,Miller:2010nz,Jaffe_PhysRevD.103.016017}.
 
Alternatively, in the infinite momentum frame (IMF), an elastic frame where $P_z \rightarrow \infty$ and $\mathbf{P}\cdot \mathbf{\Delta}=0$, a two-dimensional (2D) Fourier transform of the trace anomaly form factor in the transverse plane can be interpreted as the spatial distribution \cite{Burkardt:2000za,Miller:2010nz} and such a Fourier transform has been worked out for the EMT~\cite{Lorce:2018egm,Freese:2021czn,Panteleeva:2022uii}.
The spatial distribution of the glue trace anomaly in the IMF is obtained  as
\begin{equation}
    \rho^{\mathrm{IMF}}_{\mathrm{H}}(\mathbf{r}_{\perp}) = \int \frac{d^2 \mathbf{\Delta}_{\perp}}{(2\pi)^2} e^{-i\mathbf{\Delta}_{\perp} \cdot \mathbf{r}_{\perp}} \left. \tilde{G}_{\mathrm{H}}(Q^2)\right\vert^{P_z \rightarrow \infty}_{\mathbf{P}\cdot \mathbf{\Delta}=0},
\end{equation}
where  $\textbf{r}_{\perp}$ and $\mathbf{\Delta}_{\perp}$ are defined on the two-dimensional transverse space. In this work, for the pion and nucleon, we use the 2D Fourier transform in the IMF to obtain the spatial distributions.

\begin{figure}[!htbp]
    \centering
    \includegraphics[height=5cm,width=0.8\linewidth]{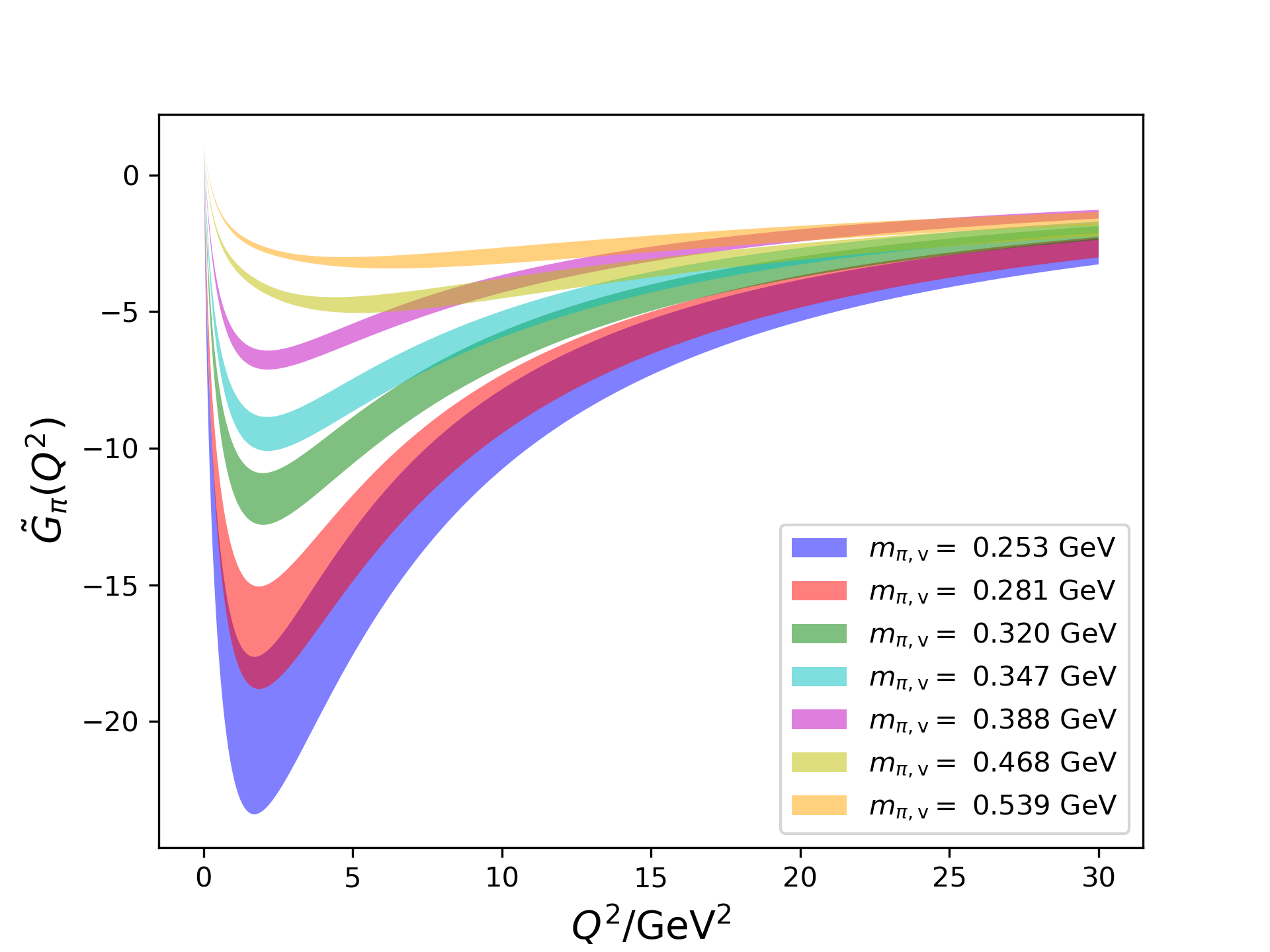}
    \includegraphics[height=5cm,width=0.8\linewidth]{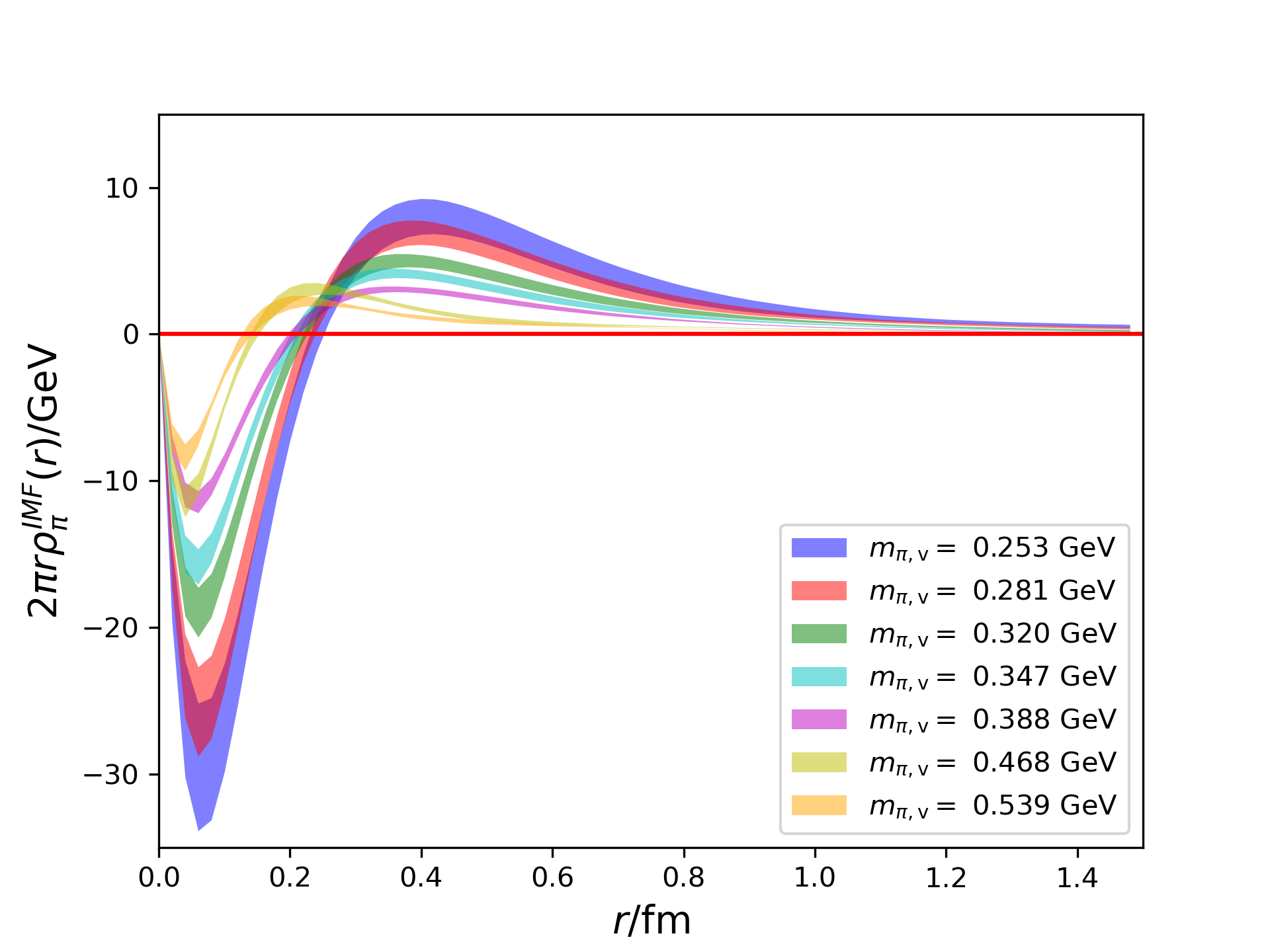}
        \caption{Top panel: the extrapolation of form factors of the pion to the large $Q^2$ region.
        Bottom panel: the glue trace anomaly spatial distribution $2\pi r\,\rho^{\mathrm{IMF}}_{\pi}(r)$ at seven valence pion masses.}
        \label{Fig:pion_trace_anomaly_density_multimass_normed_zexp_fit} 
\end{figure}

To perform the Fourier transform, we extrapolate the $z$-expansion fit  for pion glue trace anomaly form factor to large $Q^2$ where the form factors are suppressed as $1/Q^4$ due to the constraints in Eq.~(\ref{eqn:zxep_power_constrain}). The 2D spatial distributions of the glue trace anomaly in the IMF with different quark masses are shown in the bottom panel in Fig.~\ref{Fig:pion_trace_anomaly_density_multimass_normed_zexp_fit}. The sign change of the spatial distribution is observed and the intersection with the zero axis shifts to a smaller $r$ region with increasing quark mass. This is consistent with the finding in Ref.~\cite{Fangcheng_PhysRevD.104.074507}.

\begin{figure}[!htbp]
    \centering
    \includegraphics[height=5cm,width=0.8\linewidth]{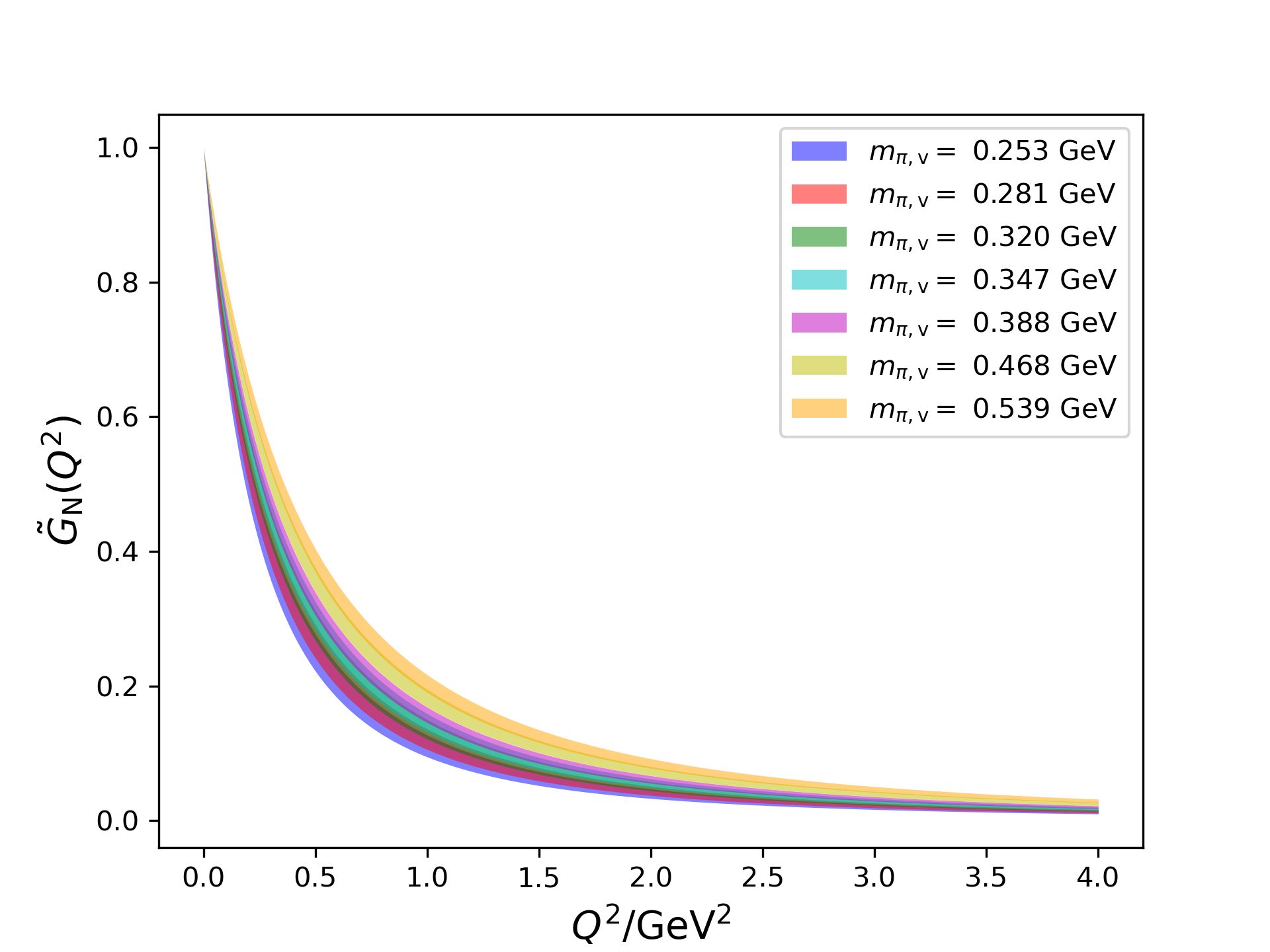}
    \includegraphics[height=5cm,width=0.8\linewidth]{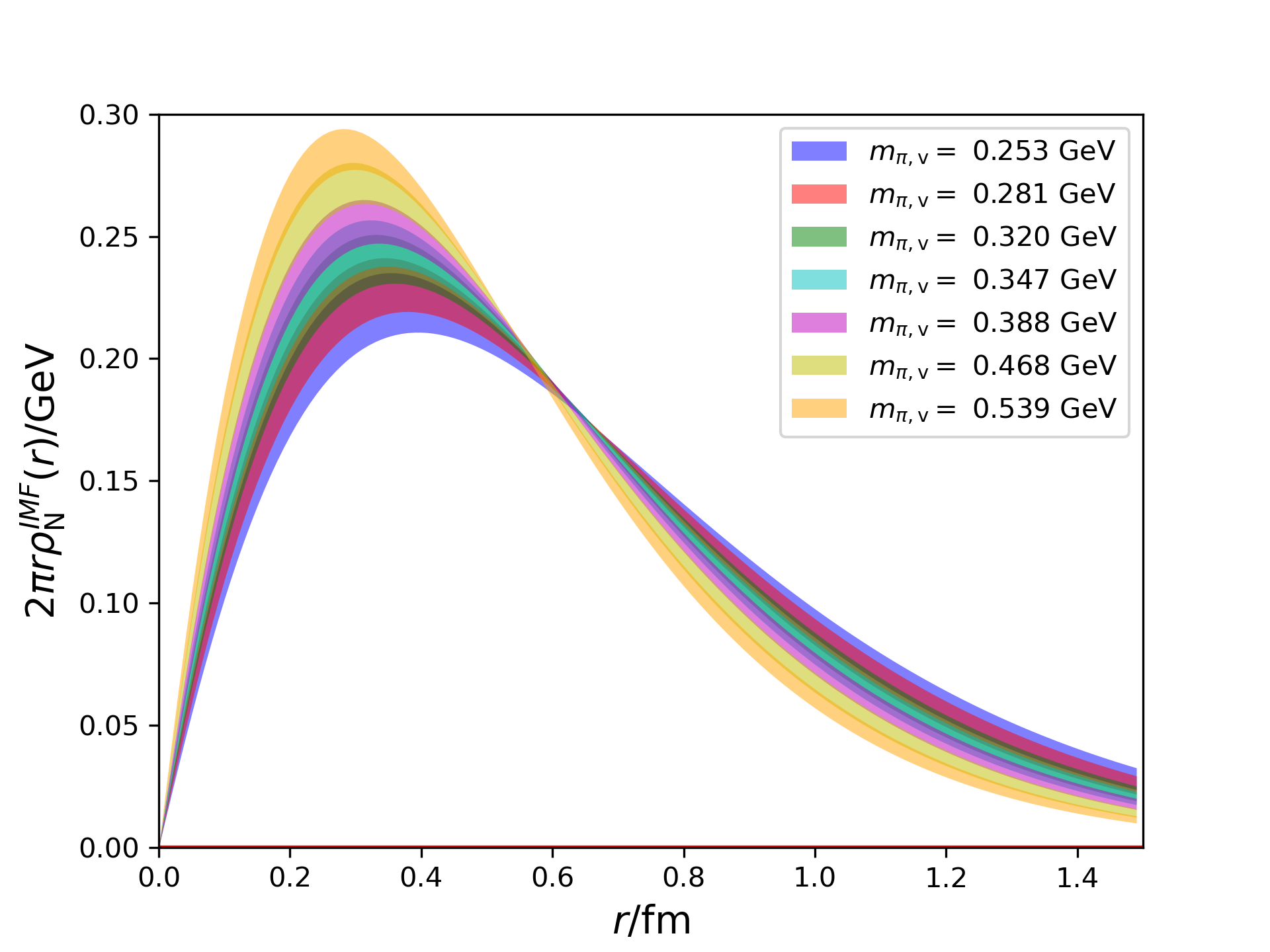}
        \caption{Top panel: the extrapolation of form factors of the nucleon to the large $Q^2$ region. Bottom panel: the glue trace anomaly spatial distribution $2\pi r\,\rho^{\mathrm{IMF}}_{N}(r)$ at seven valence pion masses.}
        \label{Fig:proton_trace_anomaly_density_multimass_normed_dipole_fit}
\end{figure}

The results for form factors and spatial distributions of the glue trace anomaly in the nucleon are shown in Fig.~\ref{Fig:proton_trace_anomaly_density_multimass_normed_dipole_fit}. In contrast to the pion case, the spatial distribution of the trace anomaly in the nucleon is always positive and the maximum value is larger with larger quark mass. The spatial distribution for the pion has a longer tail than that for the nucleon, which explains why the mass radius of the pion is larger than that of the nucleon. 


\section{Conclusion and Outlook}
\label{sec:conclusion_and_outlook} 

We have presented a calculation of the glue trace anomaly form factors of EMT and mass spatial distributions for the pion and the nucleon using overlap fermions with a range of valence pion masses on a RBC-UKQCD domain-wall ensemble.

We find that for the pion the glue trace anomaly form factor shows a sign change in the small $Q^2$ region, which is  consistent with predictions from chiral perturbation theory \cite{Novikov:1980fa, Chen_ChPT_TAFF_1998, Hatta_note_Chiral_Perturb_Soft_Pion_Small_Q2}, while for the nucleon the glue trace anomaly form factor shows no sign change.  The predictions of asymptotic signs of the trace anomaly form factors from a recent perturbative QCD calculation at large $Q^2$~\cite{Tong:2022zax} agree with this work for the pion case but disagree for the nucleon case.
We perform Fourier transforms to the form factors and obtain the spatial distribution of the glue trace anomaly in the pion and nucleon. On the one hand, we find that the spatial distribution in the pion is negative at short  distance and positive at long distance. This explains how the trace anomaly contribution to the pion mass approaches zero in the chiral limit~\cite{liu2023hadrons} and reflects the sign change in the form factor. Such a behavior in the form factor deserves to be further investigated both experimentally and theoretically. On the other hand, the spatial distribution in the nucleon is always positive and is suppressed in the large $r$ region. We also calculate the radius of the trace anomaly form factor; the radius of the pion is much larger than that of the nucleon. This observation aligns with our spatial distribution findings, as the spatial distribution within the pion exhibits an extended tail when compared to the nucleon's spatial distribution. 

Thus this work shows that the glue trace anomaly form factor can be studied accurately and efficiently with overlap fermion which preserves chiral symmetry and reveals the mass distribution within hadrons. In the future, we will include the quark contribution and extend the calculation to ensembles with physical quark masses and smaller lattice spacings for the extrapolation to the continuum limit.

\section*{Acknowledgement}
We thank the RBC/UKQCD collaborations for providing their domain-wall gauge configurations and also thank S. Brodsky, Y. Hatta, D. Kharzeev, F. Yuan, D. A. Pefkou, M. Peskin, A. Walker-Loud, R. A. Briceño, and I. Zahed for constructive discussions.

This work is supported in part by NSFC Grants No. 12293060, No. 12293062, and No. 12047503, the Strategic Priority Research Program of Chinese Academy of Sciences, Grants No.\ XDB34030303 and No. YSBR-101, and also a NSFC-DFG joint grant under Grant No.\ 12061131006 and SCHA 458/22. F. H. is supported by the National Science Foundation under Award No. PHY-1847893. J. L. is supported by NSFC under Grants No.\ 12175073 and No.\ 12222503.
G. W. is supported by the French National Research Agency under Contract No. ANR-20-CE31-0016.
This work is supported in part by the U.S.\ Department of Energy, Office of Science, Office of Nuclear Physics, under Grant No.\ DE-SC0013065. 
The authors acknowledge partial support by the U.S.\ Department of Energy, Office of Science, Office of Nuclear Physics under the umbrella of the Quark-Gluon Tomography (QGT) Topical Collaboration with Award No. DE-SC0023646. This material is based upon work partially supported by the U.S. Department of Energy, Office of Science, Office of Nuclear Physics, under Contract No. DE-AC02-05CH11231. This research used resources of the National Energy Research Scientific Computing Center (NERSC), a U.S.\ Department of Energy Office of Science User Facility located at Lawrence Berkeley National Laboratory, operated under Contract No.\ DE-AC02-05CH11231. This research used resources of the Oak Ridge Leadership Computing Facility, which is a DOE Office of Science User Facility supported under Contract No. DE-AC05-00OR22725. 
We acknowledge the facilities of the USQCD collaboration used for this research in part, which are funded by the Office of Science of the U.S.\ Department of Energy.



\bibliographystyle{apsrev4-1}
\bibliography{chiQCD_2023_Bigeng.bib}

\begin{widetext}
\section*{Appendix}
\subsection{Normalization convention of the single-particle states}
\label{app:norm_convention}
The completeness relation for the $\chi QCD$ convention is 
\begin{equation}
1=\int \frac{d^3p}{(2\pi)^3}|p\rangle\frac{m}{E_p}\langle p|,~~~|p\rangle =\sqrt{\frac{E_p}{m}}a_p^+|\Omega\rangle   
\end{equation}
The one-particle state is defined as 
\begin{equation}
\phi(x)|\Omega\rangle=\int\frac{d^3p}{(2\pi)^3}\frac{\sqrt{2m}}{2E_p}e^{-i\vec{p}\cdot\vec{x}}|p\rangle,~~~  
\langle\Omega|\phi^{\dag}(x)=\int\frac{d^3p}{(2\pi)^3}\frac{\sqrt{2m}}{2E_p}e^{i\vec{p}\cdot\vec{x}}\langle p|,
\end{equation}
\begin{equation}
\langle \Omega|\phi^{\dag}(p)|l\rangle=\int d^3xe^{-i\vec{p}\cdot\vec{x}}\langle\Omega|\phi^{\dag}(x)|l\rangle=\delta^3(\vec{p}-\vec{l})(2\pi)^3\sqrt{2m}.
\end{equation}
In this definition, $|p\rangle=\tilde{|p\rangle}/\sqrt{2m}$ where $\tilde{|p\rangle}$ is the single-particle state with the conventional normalization where
\begin{align}
    1=\int \frac{d^3p}{(2\pi)^3}\tilde{|p\rangle}\frac{1}{2E_p}\tilde{\langle p|}, ~~~\tilde{|p\rangle} =\sqrt{2E_p}a_p^+|\Omega\rangle,
\end{align}
Then we should have the relation between the matrix elements for these two conventions
\begin{align}
\langle p|O_g(0)|0\rangle = \frac{1}{2m_{\mathrm{H}}}\tilde{\langle p|}O_g(0)\tilde{|0\rangle }.
\end{align}

Therefore, the dimensionless form factor for pion is defined as
\begin{align}
 \tilde{\langle p_f|}O_g(q)\tilde{|p_i\rangle }=2m_{\mathrm{H}}^2 G_{\mathrm{H}}(t),
\end{align}
where $t=(p_f-p_i)^2$.
The form factor can be extracted from the matrix element under the $\chi QCD$ normalization convention:
\begin{align}
\langle p_f|O_g(q)|p_i\rangle  = \frac{1}{2m_{\mathrm{H}}} \tilde{\langle p_f|}O_g(q)\tilde{|p_i\rangle} =\frac{1}{2m_{\mathrm{H}}} 2m_{\mathrm{H}}^2 G_\pi(t) = m_{\mathrm{H}} G_{\mathrm{H}}(t)
\end{align}

\subsection{Two-point correlation functions}
The pion two-point correlation function with momentum $\vec{p}$ can be written as
\begin{align}
    \label{eqn:pion_2pt}
    \begin{aligned}
        C_{\pi,\mathrm{2pt}}(t;\vec{p})  = & \sum_{\vec{x}} e ^{-i \vec{p} \cdot \vec{x}} \langle \chi_{\pi}(\vec{x},t) \chi_{\pi}^{\dag}(\vec{0},0)  \rangle \\
        = & \sum_{\vec{x}} e^{-i \vec{p} \cdot \vec{x}}   
        \sum_{n',\vec{p}_2} \frac{m}{E^{n'}_{\vec{p}_2}}
        \sum_{n,\vec{p}_1}\frac{m}{E^n_{\vec{p}_1}} \langle n',\vec{p}_2 |\chi_{\pi}(\vec{x},t) | n,\vec{p}_1 \rangle \langle  n,\vec{p}_1 |\chi_{\pi}^{\dag}(\vec{0},0) | n',\vec{p}_2  \rangle\\
        & \xrightarrow[]{t \gg  0} \frac{m_{\pi}}{E_{\pi,\vec{p}}} Z^{S_i}_{\vec{p}} Z^{S_f}_{\vec{p}} \times (e^{-E t } + e^{-E(T - t)}),
    \end{aligned} 
\end{align}
where $S_i$ and $S_f$ denote the smearing settings of the source and sink.

Using the projection operators,
\begin{align}
    \Gamma_{\pm} = \frac{1}{2} \bigg ( 1 \pm  \gamma_4 \bigg ),
\end{align}
the nucleon two-point correlation function with momentum $\vec{p}$ and summed spin $s$ can be written as
\begin{align}
    \label{eqn:proton_2pt}
    \begin{aligned}
        C_{N,\mathrm{2pt}}(t;\vec{p}) &= \mathrm{Tr}[\Gamma_{+}G_{N,\mathrm{2pt}}(t;\vec{p})]\\
        & \xrightarrow[]{t \gg  0}   \frac{m_+}{E^+_{p}}  e^{-E^+_{p} t } Z^{S_i}_{\vec{p}} Z^{S_f}_{\vec{p}} \frac{\mathrm{Tr}[\frac{1}{2}(1 +  \gamma_4)(-i\slashed{p}+m_+)]}{2m_+} =  \frac{m_N + E_{N,\vec{p}}}{E_{N,\vec{p}}} Z^{S_i}_{\vec{p}} Z^{S_f}_{\vec{p}} e^{-E_{N,\vec{p}} t }.
    \end{aligned}
\end{align}

The general functional form of the 2pt functions can be written as
\begin{align}
    C_{\mathrm{H},\mathrm{2pt}}(t;\vec{p}) = \mathcal{K}_{\mathrm{H},\mathrm{2pt}}(p) Z^{S_i}_{\vec{p}} Z^{S_f}_{\vec{p}}  \mathcal{T}_{2}(t, T) + \sum_{n=1}^{n_\mathrm{max}} \sum_{\alpha_n}A^n_{\alpha}\tilde{\mathcal{T}}^n_{\mathrm{2pt}, \alpha}(t,  T),
\end{align}
where $t$ is the source-sink time separation and $\tau$ is the current-source time separation.  $\mathcal{K}_{\mathrm{H},\mathrm{2pt}}(p_i,p_f)$ is the kinematic factor, which can be derived from Eqs.~(\ref{eqn:pion_2pt}) and (\ref{eqn:proton_2pt}) and are shown in Table~\ref{tab:kinematics}. The energies appearing in these expressions can be calculated using the dispersion relationship or using fitted parameters $E_i$ and $E_f$, etc. $Z_{\vec{p}}$ is the overlap factor between the hadron state and the interpolating operator. $\mathcal{T}_{\mathrm{2pt}}(t, \tau, T)$ is the time dependence of the ground state and the second term includes contributions from the first excited state up to the $n_{\mathrm{max}}$th excited state, with $\alpha_n$ terms summed up. $\tilde{\mathcal{T}}^n_{\mathrm{2pt}, \alpha}(t, \tau, T)$ is the corresponding time time dependence and $A^n_{\alpha}$s are the corresponding weights.

\begin{table}[!htbp]
    \centering
    \begin{tabular}{|c|c|c|} \hline
            & $\pi$ & $N$ \\ \hline
        2pt & $\frac{m_{\pi}}{E_{\pi,\vec{p}}}$ & $\frac{m_N + E_{N,\vec{p}}}{E_{N,\vec{p}}}$ \\ \hline
        3pt & $\frac{m_{\pi}}{E_{\pi,\vec{p}_i}}\frac{m_{\pi}}{E_{\pi,\vec{p}_f}} $  & $\frac{1}{2E_{N,\vec{p}_i}E_{N,\vec{p}_f}} [-(\vec{p}_i \cdot \vec{p}_f - E_{N,\vec{p}_i}E_{N,\vec{p}_f})  + m_N^2 +m_N(E_{N,\vec{p}_i} + E_{N,\vec{p}_f})]$\\ \hline
        3pt back to back & $\frac{m_{\pi}}{E_{\pi,\vec{p}_i}}\frac{m_{\pi}}{E_{\pi,\vec{p}_f}} $  & $\frac{4m_N}{8E_{p_i}E_{p_f}} [2m_N + (E_{p_i} + E_{p_f})]$\\ \hline
        3pt source at rest & $\frac{m_{\pi}}{E_{\pi,\vec{p}_f}} $  & $\frac{m_N + E_{N,\vec{p}_f}}{E_{N,\vec{p}_f}} $\\ \hline
    \end{tabular}
    \caption{The kinematic factors of the two- and three-point functions for pion and nucleon.}
    \label{tab:kinematics}
\end{table}

\subsection{Three-point correlation functions and ratio}
The pion three-point correlation function can be written as
\begin{align}
    \label{eqn:pion_3pt}
    \begin{aligned}
        C_{\pi,\mathrm{3pt}}(t,\tau;\vec{p_i},\vec{p_f})  & = \sum_{\vec{x}_f,\vec{z} } e ^{-i \vec{p}_f \cdot \vec{x}_f}  e ^{i \vec{q} \cdot \vec{z}} \langle \chi_{\pi}(\vec{x}_f,t) \mathcal{O}(\vec{z},\tau)\chi_{\pi}^{\dag}(\vec{0},0)  \rangle\\
        & \xrightarrow[]{t \gg \tau \gg 0} \langle \pi | \mathcal{O} | \pi \rangle Z_{\vec{p}_i} Z_{\vec{p}_f} \frac{m_{\pi}}{E_{p_i}}\frac{m_{\pi}}{E_{p_f}} e^{-E_i \tau - E_f(t-\tau)}
    \end{aligned} 
\end{align}
where $\vec{p}_i$ and $\vec{p}_f$ are the source momentum and sink momentum and $\vec{q} = \vec{p}_f-\vec{p}_i$ is the three-momentum injected by the glue trace anomaly operator $\mathcal{O}$.

Similarly, for the nucleon with positive parity, i.e., the proton, we have
\begin{align}
    \label{eqn:proton_3pt}
    \begin{aligned}
        C_{N^+,\mathrm{3pt}}(t,\tau;\vec{p_i},\vec{p_f})  & = \sum_{\vec{x}_f,\vec{z} } e ^{-i \vec{p}_f \cdot \vec{x}_f}  e ^{i \vec{q} \cdot \vec{z}} \mathrm{Tr}[\Gamma_+\langle \chi_{N^+}(\vec{x}_f,t) \mathcal{O}(\vec{z},\tau)\chi_{N^+}^{\dag}(\vec{0},0)  \rangle]\\
        & \xrightarrow[]{t \gg \tau \gg 0} m_{N^+} G_{N^+} Z_{\vec{p}_i} Z_{\vec{p}_f} \frac{m_{+}}{E_{p_i}}\frac{m_{+}}{E_{p_f}} e^{-E_i \tau - E_f(t-\tau)}\\
        &~~~~~~~~~~~~~~~~~\frac{1}{8m_+^2}[-4(\vec{p}_i \cdot \vec{p}_f - E_{p_i}E_{p_f})  + 4m_+^2 +4m_+(E_{p_i} + E_{p_f})], 
    \end{aligned} 
\end{align}
where we have used $\slashed{p}_i = \vec{p}_i \cdot \gamma_i + i E_{i}\gamma_4$.

The general functional form of the 3pt can be written as
\begin{align}
    C_{\mathrm{H},\mathrm{3pt}}(t,\tau) = m_{\mathrm{H}} G_{\mathrm{H}} \mathcal{K}_{\mathrm{H},\mathrm{3pt}}(p_i,p_f) Z_{\vec{p}_i} Z_{\vec{p}_f} \mathcal{T}_{\mathrm{3pt}}(t, \tau, T) + \sum_{n=1}^{n_\mathrm{max}} \sum_{\alpha_n}C^n_{\alpha}\tilde{\mathcal{T}}^n_{\mathrm{3pt}, \alpha}(t, \tau, T),
\end{align}
where $t$ is the source-sink time separation and $\tau$ is the current-source time separation.  $\mathcal{K}_{\mathrm{H},\mathrm{3pt}}(p_i,p_f)$ is the kinematic factor, which can be calculated using the dispersion relationship or using fitted parameters $E_i$ and $E_f$, etc. The kinematic factors for momentum transfer scenarios are listed in Table~\ref{tab:kinematics}. $Z_{\vec{p}_i}$ and $Z_{\vec{p}_f}$ are overlap factors between the hadron state and the interpolating operator. $\mathcal{T}_{\mathrm{3pt}}(t, \tau, T)$ is the time dependence of the ground state and the second term includes contributions from the first excited state up to the $n_{\mathrm{max}}$th excited state, with $\alpha_n$ terms summed up. $\tilde{\mathcal{T}}^n_{\mathrm{3pt}, \alpha}(t, \tau, T)$ is the corresponding time time dependence and $C^n_{\alpha}$s are the corresponding weights.

In this work, we include the first excited states and have:
\begin{align}
    \mathcal{T}_{3}(t, \tau, T) = e^{-E_i \tau - E_f(t-\tau)}
\end{align}
\begin{align}
    \tilde{\mathcal{T}}^1_{\mathrm{3pt},a}( \tau;t,T) = e^{-E^1_i \tau - E_f(t-\tau)}, ~~~
    \tilde{\mathcal{T}}^1_{\mathrm{3pt},b}( \tau;t,T) = e^{-E_i \tau - E^1_f(t-\tau)}, ~~~
    \tilde{\mathcal{T}}^1_{\mathrm{3pt},c}( \tau;t,T) = e^{-E^1_i \tau - E^1_f(t-\tau)}.
\end{align}
The functional form we used for fitting 3pt is
\begin{align}
    \begin{aligned}
        C_{\mathrm{H},\mathrm{3pt}}(t,\tau) = & m_{\mathrm{H}} G_{\mathrm{H}} \mathcal{K}_{\mathrm{H},\mathrm{3pt}}(p_i,p_f) Z_{\vec{p}_i} Z_{\vec{p}_f} e^{-E_i \tau - E_f(t-\tau)} \\ 
        &+ C_1 e^{-E^1_i \tau - E_f(t-\tau)} + C_2 e^{-E_i \tau - E^1_f(t-\tau)} + C_3 e^{-E^1_i \tau - E^1_f(t-\tau)}.
    \end{aligned}
\end{align}

\end{widetext}
\end{document}